\documentclass[
reprint,
superscriptaddress,
 amsmath,amssymb,
 aps,
floatfix,
nofootinbib
]{revtex4-1}

\usepackage[utf8]{inputenc}
\usepackage[english]{babel}

\usepackage{makecell}
\usepackage{textcomp}
\usepackage{amsmath,etoolbox}
\usepackage{amsthm}
\usepackage{amssymb}
\usepackage[left=2.5cm,right=2.5cm,top=3cm,bottom=3cm,bindingoffset=0cm]{geometry}
\usepackage{dsfont}
\usepackage{bm}
\usepackage{leftidx}
\usepackage{float}
\usepackage[makeroom]{cancel}
\usepackage{indentfirst}
\usepackage{underoverlap}
\usepackage{caption}
\usepackage{hyperref}
\usepackage{tikz}
\usetikzlibrary{quantikz}
\usepackage{subfig}
\usepackage{graphicx}
\usepackage{tikz-cd}
\usepackage[titletoc,title]{appendix}

\usepackage{adjustbox}
\usepackage{mathtools}
\usepackage{multirow}
\usepackage{xargs}
\usepackage{xstring}
\usepackage{rotating}

\usetikzlibrary{matrix,arrows,decorations.pathmorphing}
\newenvironment{sloppypar*}{\sloppy\ignorespaces}{\par}
\allowdisplaybreaks 

\newcommand {\sket} [1] {| #1 \rangle}
\newcommand {\sbraket} [2] {\langle #1 | #2 \rangle}
\newcommand {\saxe} [2] {| #1 \rangle\langle #2 |}
\newcommand {\sand} [3] {\langle #1 | #2 | #3 \rangle}

\DeclareMathOperator{\tr}{tr\s{}  }

\newcommand{\bma} {\begin{pmatrix}}
\newcommand{\ema} {\end{pmatrix}}

\newcommand{\iu}{{i\mkern1mu}}

\renewcommand{\d}[1]{\ensuremath{\operatorname{d}\!{#1}}}
\newcommand{\me}{\mathrm{e}}
\newcommand{\s}{\hspace{.08em}}

\newcommandx*\hilbdimK[1][1=]{\mathcal{D}_{K\IfEqCase{#1}{{}{}}[=#1]}}
\newcommandx*\hilbdimKQ[2][1=,2=]{\mathcal{D}_{K\IfEqCase{#1}{{}{}}[=#1],Q\IfEqCase{#2}{{}{}}[=#2]}}

\newcommand{\poly}{\operatorname{poly}}

\newcommand*{\vv}[1]{\vec{\mkern0mu#1}}

\newcommand{\OPa}{a^{\vphantom{\dagger}}}
\newcommand{\OPad}{a^{\dagger}}

\newcommandx*\bound[3][,2=,3=]{#1_{K
\IfEqCase{#2}{{}{}}[,\,#2]}^{
\IfEqCase{#3}{{}{}}[(#3)]
}}

\DeclareMathAlphabet{\mathdutchcal}{U}{dutchcal}{m}{n}

\def\ket#1{\left\vert #1 \right\rangle}
\def\bra#1{\left\langle #1 \right\vert}

\newcommand{\be}{\begin{equation}}
\newcommand{\ee}{\end{equation}}
\newcommand{\bp}{\begin{pmatrix}}
\newcommand{\ep}{\end{pmatrix}}
\newcommand{\ben}{\begin{enumerate}}
\newcommand{\een}{\end{enumerate}}

\def\mystrut(#1,#2){\vrule height #1pt depth #2pt width 0pt}

\let\oldFootnote\footnote
\newcommand\nextToken\relax
\renewcommand\footnote[1]{\oldFootnote{#1}\futurelet\nextToken\isFootnote}
\newcommand\isFootnote{\ifx\footnote\nextToken\textsuperscript{,}\fi}

\usepackage[shortcuts]{extdash}

\newcommand*\standardbin{+}

\makeatletter

\newcommand{\doublewidetilde}[1]{{%
  \mathpalette\double@widetilde{#1}%
}}
\newcommand{\double@widetilde}[2]{%
  \sbox\z@{$\m@th#1\widetilde{#2}$}%
  \ht\z@=.9\ht\z@
  \widetilde{\box\z@}%
}

\patchcmd{\env@cases}
  {\quad}
  {{ \ }}
  {}{}

\newcommand*\tabularbin[1]{%
  \mathbin{\mathpalette{\@tabularsym\standardbin}{#1}}%
}

\newcommand*\@tabularsym[3]{%
  \setbox\z@\hbox{$#2#1\m@th$}%
  \hbox to\wd\z@{\hss$#2#3\m@th$\hss}%
}

\newcommand*{\myrulefill}[3][]{%
  \makebox[#2]{#1%
    \leaders\hrule height \dimexpr.5ex+.2pt\relax depth \dimexpr -.5ex+.2pt\relax \hfill
    \enskip{#3}\enskip
    \leaders\hrule height \dimexpr.5ex+.2pt\relax depth \dimexpr -.5ex+.2pt\relax \hfill\kern0pt}
}

\usetikzlibrary{arrows,positioning}
\tikzset{
    >=stealth',
    punkt/.style={
           rectangle,
           rounded corners,
           draw=black, very thick,
           text width=6.5em,
           minimum height=2em,
           text centered},
    pil/.style={
           ->,
           thick,
           shorten <=2pt,
           shorten >=2pt,}
}

\renewcommand\@makecaption[2]{%
  \par
  \vskip\abovecaptionskip
  \begingroup
   \small\rmfamily
    \begingroup
     \samepage
     \flushing
     \let\footnote\@footnotemark@gobble
     \@make@capt@title{#1}{#2}\par
    \endgroup
  \endgroup
  \vskip\belowcaptionskip
}

\makeatother

\begin{document}

\title{Simulating Hadronic Physics on NISQ devices using Basis Light-Front Quantization}

\author{Michael Kreshchuk}
\affiliation{Department of Physics and Astronomy, Tufts University, Medford, MA 02155, USA}
\author{Shaoyang Jia}
\affiliation{Department of Physics and Astronomy, Iowa State University, Ames, IA 50011, USA}
\affiliation{Physics Division, Argonne National Laboratory, Argonne, IL 60439, USA}
\author{William M. Kirby}
\affiliation{Department of Physics and Astronomy, Tufts University, Medford, MA 02155, USA}
\author{Gary Goldstein}
\affiliation{Department of Physics and Astronomy, Tufts University, Medford, MA 02155, USA}
\author{James P. Vary}
\affiliation{Department of Physics and Astronomy, Iowa State University, Ames, IA 50011, USA}
\author{Peter J. Love}
\affiliation{Department of Physics and Astronomy, Tufts University, Medford, MA 02155, USA}
\affiliation{Computational Science Initiative, Brookhaven National Laboratory, Upton, NY 11973, USA }


\begin{abstract}
    The analogy between quantum chemistry and light-front quantum field theory, first noted by Kenneth G. Wilson, serves as motivation to develop light-front quantum simulation of quantum field theory. We demonstrate how calculations of hadron structure can be performed on Noisy Intermediate-Scale Quantum devices within the Basis Light-Front Quantization framework.
    We calculate the light-front wave functions of pions using an effective light-front Hamiltonian in a basis representation on a current quantum processor.
    We use the Variational Quantum Eigensolver to find the ground state energy and wave function, which is subsequently used to calculate pion mass radius, decay constant, elastic form factor, and charge radius.
\end{abstract}

\maketitle

\tableofcontents

\section{Introduction}

Quantum simulation of quantum field theory (QFT) is a promising application of quantum computing that has recently seen a surge in interest~\cite{feynman1982,lloyd1996universal,zalka1998simulating,wiesner1996simulations,boghosian1997quantum,meyer1996quantum,aspuru2005simulated,wu2002polynomial,preskill1,preskill2,NSFWSR,BESReport,ASCRReport,berry2017exponential,low2017optimal,low2016hamiltonian,berry2015hamiltonian,berry2015simulating,wiese2014towards,zohar2015quantum,zohar2013cold,zohar2013quantum,gonzalez2017quantum,zhang2016fermion,opportunities,barrett2013simulating, marshall2015quantum,2017arXiv171104006H,bauer2019,martinez2016real,muschik2017u,Liu:2020eoa}.
In our previous work~\cite{Kreshchuk:2020dla} we demonstrated that the light-front (LF) quantization of quantum field theory provides a natural framework for \emph{ab initio} digital quantum simulation.
The Discretized Light-Cone Quantization (DLCQ) technique allows one to significantly reduce the quantum-computational resources when compared to lattice approaches.

In the present paper, we continue our program of investigating quantum simulation in the light-front (LF) formulation, and develop an approach to simulating field theory based on the Basis Light-Front Quantization~(BLFQ)~\cite{varybasis,Zhao:2014hpa} technique.
The basis-function expansion allows us to further reduce the need for computational resources, making calculations accessible for  existing quantum devices.

In~\cite{Kreshchuk:2020dla} we developed quantum algorithms based on simulating time evolution and adiabatic state preparation. In this work, we instead aim for near-term devices by adopting the Variational Quantum Eigensolver (VQE) paradigm: this allows us to implement a demonstration on the IBM Vigo quantum processor. VQE is a hybrid quantum-classical algorithm. The classical computer optimizes the expectation value of the Hamiltonian, which is repeatedly evaluated on a quantum computer. The classical computer/algorithm optimizes the parameters of this ansatz to minimize the ground state energy.

The resulting parametrized quantum circuit approximately prepares the ground state wave function of the Hamiltonian.
Thus once the VQE procedure is complete, we can compute the expectation values of other observables in this approximate ground state.

The DLCQ and BLFQ paradigms provide alternative approaches to describing relativistic interactions. While both are, in principle, \emph{ab initio} frameworks, DLCQ studies the system starting from the light-front Hamiltonian quantized in the traditional free-field basis placed on a discrete momentum grid. BLFQ starts from the same light-front Hamiltonian but quantizes it in terms of modes tailored to the  symmetries of the system under consideration in order to construct an computationally efficient representation of the Hamiltonian. Since each represents a choice of basis spaces for the fields, they should yield the same results in fully converged calculations, \emph{i.e.}, in their respective continuum limits.

Having much in common with \emph{ab initio} quantum chemistry and nuclear theory, the BLFQ formulation provides an ideal framework for benchmarking NISQ devices and testing existing algorithms on physically relevant problems such as the calculation of hadronic spectra~\cite{LI2016118,Li:2017mlw,Tang:2018myz,Tang:2019gvn} and parton distribution functions (PDFs)~\cite{Lan:2019vui,Lan:2019img,Lan:2019rba}.
In essence, BLFQ amounts to (1) Choosing the effective field theory most efficiently describing the problem of interest, (2) Quantizing the system in the light-cone coordinates, (3) Non-perturbatively solving the theory in the most suitable basis. This results in an efficient representation of the QFT problem under study. One typically starts with a fixed-particle-number formulation, effectively reducing the QFT setting to a relativistic quantum-mechanical many-body problem. In many cases, already at this level one can obtain results with suitable precision to make meaningful comparisons with experimental results~\cite{Li:2017mlw,Tang:2018myz,Tang:2019gvn,basislightmesons,Lan:2019vui,Lan:2019rba,Lan:2019img}.

In this article, we consider the dynamics of valence quarks for light mesons on the light front using the Hamiltonian from~\cite{basislightmesons}. This Hamiltonian includes the kinetic energy, the confinement potential in both the longditudinal and the transverse directions~\cite{Li:2017mlw}, and the Nambu--Jona-Lasinio interaction~\cite{Klevansky:1992qe} to account for the chiral interactions among quarks. We limit ourselves to the valence Fock sector of mesons while working with relative momentum variables. The dependence of the light-front wave functions for these valence quarks on the relative momentum is expanded in terms of orthonormal basis functions. After implementing finite cut-offs in this expansion, the light-front Hamiltonian becomes a hermitian matrix in the basis representation. We use the same scheme as in Ref.~\cite{basislightmesons} to fix our model parameters at each choice of basis cut-offs. We illustrate our ideas by choosing an effective Hamiltonian for the light meson system, and running the VQE minimization on the IBM Vigo machine to calculate the squared pion mass. Using the resulting wave function, we calculate squared mass, decay constant,  mass radius, electromagnetic form factor, and charge radius of the pion.

The two different ans\"atze we consider in this paper are based on different ways of encoding physical states on the quantum computer. Within the \emph{direct encoding} one stores the occupancies of the second-quantized states in the unary form and uses the Unitary Coupled Cluster ansatz for state preparation. Within a more efficient \emph{compact encoding}, one stores the occupancies in the binary form, which requires logarithmically fewer qubits and allows one to prepare an ansatz state using the arbitrary state preparation algorithms, given that the particle number is fixed and small.

In Sec.~\ref{sec:blfq}, we provide a summary of the BLFQ formalism and a representation of basis functions, which we used throughout the paper. In Sec.~\ref{sec:observ}, we derive expressions for various observables in the chosen basis. In Sec.~\ref{ss:qc_methods} we describe two variations of the VQE algorithm, and show the results of running it on an existing quantum computer.

\section{Basis Light-Front Quantization \label{sec:blfq}}

\subsection{Overview\label{sec:lfoverview}}
	The light-front quantization approach specifies the commutation relation of fields at equal light-front time~\cite{Brodsky:1997de}. In contrast to the Lagrangian formulation of equal-time quantization, the field theory dynamics after light-front quantization is governed by a light-front Hamiltonian~\cite{Brodsky:1997de} responsible for the light-front time evolution of the system. Quantizing a QFT on the light front has a number of advantages: triviality of the vacuum, absence of ghost fields in the light-cone gauge, Hamiltonian sparsity, and the simple form of observables in terms of the wave functions.	
	Within the light-front approach, the  bound state masses and associated wave functions are solvable from the light-front-time-independent Schr\"{o}dinger equation.
	
	In~\cite{Kreshchuk:2020dla} we developed a simulation algorithm based on the DLCQ, allowing one to reduce the resource requirements for the \emph{ab initio} simulation of QFT by a few orders of magnitude. In this work, we further reduce the computational requirements into the range of the capabilities of existing quantum devices.
	We achieve this by employing the framework of BLFQ~\cite{varybasis,Zhao:2014hpa}.
	
	Within BLFQ, a field is expanded in terms of second-quantized Fock states representing occupancies of modes (first-quantized basis functions), and there is no \emph{a priori} limit on the degrees of freedom~\cite{varybasis,Vary:2016ccz}. Accordingly, our algorithms are designed to efficiently simulate  QFT applications where particle number is not conserved.
However, for QFTs at low resolution, BLFQ is often restricted to the valence degrees of freedom, allowing the adoption of this restriction in order to implement quantum simulations on an existing quantum chip.
These experiments represent the first stage shown in Fig.~\ref{tab:roadmap}, which illustrates a progression of methods that scale towards fault-tolerant simulation of
QFTs in the quantum supremacy regime.
However, the methods we propose apply to the first three stages in Tab.~\ref{tab:roadmap}. The final stage was discussed in~\cite{Kreshchuk:2020dla}.

	Previous development of BLFQ for the heavy mesons is partially based on the holographic confinement potential between the valence quark and antiquark in the holographic transverse directions~\cite{Li:2017mlw,Tang:2018myz,Tang:2019gvn}. This potential is supplemented by a longitudinal confinement potential to attain a 3-dimensional spherical confinement potential in the nonrelativistic limit. These potentials are constructed independent of the spins for the quark and the antiquark and they are governed by a single overall strength parameter. In addition to the kinetic energy and the confinement potentials, they form the baseline Hamiltonian that is analytically solvable and defines our basis
	functions~\cite{Li:2017mlw,basislightmesons}. These basis functions possess desired spatial symmetries and boost invariances. The derived effective one-gluon exchange interaction based on the gauge dynamics serves as the spin-orbit interaction and incorporates a running coupling~\cite{Li:2017mlw}.
	
	In this article we adopt the Hamiltonian in Ref.~\cite{basislightmesons} for the light mesons. Specifically the same confinement potential forms as those in Ref.~\cite{Li:2017mlw} are implemented. However, we do not include the one-gluon exchange because the interactions for light quarks manifest from the chiral symmetry, which is insufficiently accounted for by a perturbative expansion of the gauge interaction. Instead, we resort to the Nambu--Jona-Lasinio (NJL) model for the chiral interaction of these quarks~\cite{Vogl:1989ea,Vogl:1991qt,Klevansky:1992qe}. Within our basis representation, the matrix elements of the NJL interaction can be calculated analytically~\cite{basislightmesons}. We compute the lowest mass eigenvalue and its corresponding eigenvector, its light-front wave function, using the algorithm to be described in Section~\ref{ss:qc_methods}. We then calculate observables based on this eigenvector.

\begin{table*}[htp]
\renewcommand{\arraystretch}{1.2}
\begin{tabular}{|l|l|l|l|l|}
\hline
\multirow{4}{*}{\textbf{Regime}} & \multicolumn{3}{c|}{\textbf{VQE}} & \textbf{Fault-tolerant}  \\ \cline{2-5}
 & \makecell[l]{\bf Two-body sector\\\bf BLFQ, relative\\\bf coordinate basis} &   \makecell[l]{\bf Valence sector\\\bf BLFQ, single-\\\bf coordinate basis}        &   \makecell[l]{\bf Multi-particle\\\bf BLFQ, single-\\\bf coordinate basis}   & \makecell[l]{\bf Multi-particle\\\bf DLCQ, single-\\\bf coordinate basis}  \\ \hline
Encoding & Compact & \multicolumn{2}{l|}{Compact / Direct} & Compact \\ \hline
State preparation & \makecell[l]{Arbitrary state\\preparation} & \multicolumn{2}{l|}{\makecell[l]{Arbitrary state preparation /\\Unitary Coupled Cluster / QITE}}      & \makecell[l]{Adiabatic\\state preparation}  \\ \hline
Measurement & Pauli &   \multicolumn{2}{l|}{Pauli / Sparse}      & Sparse  \\ \hline
\end{tabular}
\caption{
Flow of growing complexity and computational resources (left to right) for quantum simulation of quantum field theory on the light-front. Basis light-front quantization (BLFQ) may be considered to encapsulate discrete light-cone quantization (DLCQ).
However, we use the distinct terms here to emphasize that classical preprocessing is used in BLFQ with minimal bases for the purpose of obtaining approximations using relatively few quantum resources. The goal is to accelerate convergence to the continuum limit for bound state observables and, hence, to optimally use existing quantum resources in the NISQ era for these problems.  Treatment of open systems, such as resonances and strong decays, will likely require DLCQ to be implemented on future fault-tolerant quantum computers.
\label{tab:roadmap}
}
\end{table*}

	\subsection{The effective Hamiltonian of the BLFQ-NJL model\label{sec:blfq-njl}}

    The light-front wave functions (LFWFs) of the valence quarks for the $\pi^+$ meson and the $K^+$ meson have been solved from Ref.~\cite{basislightmesons} in the basis light-front quantization (BLFQ) framework using Nambu--Jona-Lasinio interactions~\cite{Klimt:1989pm,Vogl:1989ea,Vogl:1991qt,Klevansky:1992qe} on a classical computer. Specifically, one first truncates the light-front wave-function for the mesons to the valence quark Fock sector such that the state vector is expressed as
	\begin{equation}
	\label{eq:Psi_meson_qqbar}
	\begin{alignedat}{9}
&\bigl|\Psi(P^+,\vv{P}^\perp)\bigr\rangle
=\sum_{r,s}\int_{0}^{1}\dfrac{dx}{4\pi x(1-x)}
\\&
\hspace{-.05cm}\times\int\dfrac{d\vv{\kappa}^\perp}{(2\pi)^2}\,\psi_{rs}(x,\vv{\kappa}^\perp)
\times b_r^\dagger(xP^+,\vv{\kappa}^\perp+x\vv{P}^\perp)
\\& \hspace{.81cm}
\times d_s^\dagger((1-x)P^+,-\vv{\kappa}^\perp+(1-x)\vv{P}^\perp)\sket{0} \ .
\end{alignedat}
	\end{equation}
	where $P=k+p$ is the total momentum of the meson, $x=k^+/P^+$ is the longitudinal momentum fraction carried by the valence quark, and $\vv{\kappa}^\perp=\vv{k}^\perp-x\vv{P}^\perp$ is the relative transverse momentum.
	
	In order to solve for the LFWFs for the valence quarks inside light mesons, we adopt the effective Hamiltonian that can be represented as a basis-diagonal term and the NJL interaction:
	\begin{equation}\label{eq:decompose_Heff}
    H_{\mathrm{eff}}=H_0+H^{\mathrm{eff}}_{\mathrm{int}} \ .
    \end{equation}
	The basis-diagonal term $H_0$ contains the kinetic energy of the valence quarks, the transverse confinement potential, and the longitudinal confinement potential. In the valence Fock sector of mesons, this term takes the form of
	\begin{equation}
	\label{eq:H0_def}
	\begin{alignedat}{8}
&H_0 =\dfrac{(\vv{\kappa}^\perp)^2+\mathbf{m}^2}{x}+\dfrac{(\vv{\kappa}^\perp)^2+\overline{\mathbf{m}}^2}{1-x}
\\&+b^4x(1-x)\vv{r}_\perp^2
    -\dfrac{b^4}{(\mathbf{m}+\overline{\mathbf{m}})^2}\partial_xx(1-x)\partial_x \ ,
\end{alignedat}
	\end{equation}
	where $x$ is the longitudinal momentum fraction carried by the valence quark and $\vv{\kappa}_\perp$ is the relative transverse momentum of the valence quarks. The masses of the valence quark and the valence antiquark are given by $\mathbf{m}$ and $\overline{\mathbf{m}}$, respectively. In a addition, $b$ specifies the strength of the confinment potentials. This part of the Hamiltonian has analytic solutions that constitute the basis states for the BLFQ approach as will be seen in detail in Subsection~\ref{ss:basis_exp}.
	
	When quarks in the confinement region are the retained degrees of freedom, the strong interaction among them can be understood to arise from the global chiral symmetry, an approximate symmetry of quantum chromodynamics. To model this chiral interaction, we employ the interaction in the scalar-pseudoscalar channel of the color-singlet NJL model~\cite{Klevansky:1992qe}. Specifically, we ignore both the instantaneous interaction and the self-energy correction from the NJL interaction to obtain the following term in the total Hamiltonian:
	\begin{equation}
	\label{eq:H_eff_NJL_pi_ori}
	\begin{alignedat}{8}
H^{\mathrm{eff}}_{\mathrm{int}} = H_{\mathrm{NJL},\pi}^{\mathrm{eff}}
    =&\int \d{x}^-\int\d{}\vv{x}^\perp\,\bigl(-\dfrac{G_\pi P^+}{2}\bigr)
    \\&\times\left[\left(\overline{\psi}\psi \right)^2+\left(\overline{\psi}i\gamma_5\vv{\tau}\psi \right)^2 \right] \ ,
\end{alignedat}
	\end{equation}
    Here $\psi$ is the fermion field operator, $G_\pi$ is the NJL coupling constant, and $P^+$ is the total light-front longitudinal momentum of the system. We then expand eq.~\eqref{eq:H_eff_NJL_pi_ori} into relevant combinations of ladder operators for the quark fields. In the basis representation, this term further takes the form of a hermitian matrix, the elements of which can be calculated analytically~\cite{basislightmesons}.
	
	In this work, we solve the eigenvalue problem defined by eq.~\eqref{eq:decompose_Heff} in the total angular momentum $J_z=0$ block with the lowest eigenstates of $H_0$ forming the longitudinal and radial basis states for the interacting Hamiltonian.
	In this representation, the effective Hamiltonian takes the form of a $4$-by-$4$ matrix indexed by the basis quantum number $\theta$ that specifies the angular and spin excitations. The explicit expressions for elements in this matrix are given in Appendix~\ref{ss:H_eff_4_by_4}.
	\subsection{The basis function representations of wave functions for valence quarks of mesons\label{ss:basis_exp}}
	We adopt the following expansion of the light-front wave function for the valence quarks given by eq.~\eqref{eq:Psi_meson_qqbar}:
	\begin{align}
	\begin{alignedat}{8}
	 \psi_{rs}&(x,\vv{\kappa}^\perp) \\
     & =  \sum_{nml} \psi_{nmlrs}\, \phi_{nm}\left(\dfrac{\vv{\kappa}^\perp}{\sqrt{x(1-x)}};b\right)
    \chi_l(x)
    \ ,
	\end{alignedat}
    \label{eq:psi_rs_basis_expansions}
	\end{align}
	where $\psi_{nmlrs}$ is the expansion coefficient, $\phi_{nm}$ is a 2-dimensional (2D) harmonic oscillator (HO) eigenfunction, and $\chi_l$ is the longitudinal basis function. Here $r$ and $s$ are the spin indices of the quark and the anti-quark. Each term in eq.~\eqref{eq:psi_rs_basis_expansions} is an eigenfunction of $H_0$ in eq.~\eqref{eq:H0_def}. Explicitly, $\phi_{nm}$ is defined as
    \begin{equation}\label{eq:def_phi_nm}
    \begin{alignedat}{8}
	&\phi_{nm}\left(\vv{q}^\perp;b \right) =\dfrac{1}{b}\sqrt{\dfrac{4\pi n!}{(n+|m|)!}} \left(\dfrac{\vert\vv{q}^\perp\vert}{b}\right)^{|m|}
	\\&\times \exp\left(-\dfrac{\vv{q}^{\perp 2}}{2b^2}\right)
	\times \,L_n^{|m|} \left(\dfrac{\vv{q}^{\perp 2}}{b^2}\right)\,\exp^{im\varphi} \ ,
	\end{alignedat}
	\end{equation}
	with $\tan(\varphi)=q^2/q^1$ and $L_n^{|m|}$ being the associated Laguerre function. The parameter $b$ sets the scale of the harmonic oscillator eigenfunction, which we choose to be identical to the confining strength in the light-front Hamiltonian. Meanwhile, $\chi_l(x)$ is given by
	\begin{equation}\label{eq:def_chi_l}
    \begin{alignedat}{8}
        \quad \chi_l(x;\alpha,\beta)
	&= \sqrt{4\pi(2l+\alpha+\beta+1)}
	\\&\times\sqrt{\dfrac{\Gamma(l+1)\Gamma(l+\alpha+\beta+1)}{\Gamma(l+\alpha+1)\Gamma(l+\beta+1)}}
	\\
	& \times x^{\beta/2}(1-x)^{\alpha/2}\,P_l^{(\alpha,\beta)}(2x-1) \ ,
    \end{alignedat}
	\end{equation}
	with $P_{l}^{(\alpha,\beta)}(z)$ being the Jacobi polynomial and
	\begin{subequations}\label{eq:def_alpha_beta}
		\begin{align}
		\alpha& =2\overline{\mathbf{m}}(\mathbf{m}+\overline{\mathbf{m}})/\kappa^2 \ ,\\
		\beta&=2\mathbf{m}(\mathbf{m}+\overline{\mathbf{m}})/\kappa^2 \ .
		\end{align}
	\end{subequations}
		
	When we solve the eigenvalue problem defined by the BLFQ-NJL Hamiltonian, the following cutoffs on the basis quantum numbers following Ref.~\cite{basislightmesons} are imposed:
	\begin{equation}
	\begin{cases}
	0 \leq n \leq N_{\mathrm{max}} \\
	-M_{\mathrm{max}} \leq m \leq M_{\mathrm{max}} \\
	0 \leq l \leq L_{\mathrm{max}}
	\end{cases} \ .
	\end{equation}
	Because truncations on different basis quantum numbers are independent, we call this truncation scheme the orthogonal enumeration. Such a scheme allows us to solve simultaneously for eigenstates with different azimuthal angular momentum projection $J_z$ since it is a good quantum number in this basis. The size of the Hamiltonian in the basis representation with this orthogonal enumeration is $n_{\mathrm{H}}$-by-$n_{\mathrm{H}}$, with
	\begin{equation}\label{eq:n_H_orthogonal}
	n_{\mathrm{H}} = 4(N_{\mathrm{max}}+1)(2M_{\mathrm{max}}+1)(L_{\mathrm{max}}+1) \ .
	\end{equation}
	
	However, the capacity of NISQ devices motivates further reduction in the dimension of the Hilbert space spanned by our basis representation. Specifically, because eigenfunctions of this Hamiltonian have fixed azimuthal angular momentum projection $J_z$, the basis quantum number $\theta$ indexes specific combinations of the spin and orbital bases in the orthogonal enumeration as specified in Appendix~\ref{ss:ut_fixed_Jz}. Each basis state in the fixed $J_z$ block is then given by the basis quantum numbers $n$, $l$, and $\theta$. In the limit of $M_{\mathrm{max}}=2$, the unitary transformation that relates the bases in the fixed $J_z$ blocks to those in the orthogonal enumeration is given by Table~\ref{tab:def_Jz_theta}. The degeneracy in the basis quantum number $\theta$ in each $J_z$ block is apparent in Table~\ref{tab:def_Jz_theta}. For example when $J_z=0$, this degeneracy is $d_\theta = 4$. With a given set of $(n,\,l,\,\theta)$ in a given $J_z$ block, we take the convention such that the index of this basis is given by
	\begin{equation}
	a(n,l,\theta)= [n\,(L_{\mathrm{max}}+1)+l]\,d_\theta+\theta \ .
	\end{equation}
	For a given index, the corresponding basis quantum numbers can be easily calculated. Consequently, the size of the Hamiltonian for a fixed $J_z$ in this new enumeration becomes
	\begin{equation}
	n_{\mathrm{H}0}=d_\theta (N_{\mathrm{max}}+1)(L_{\mathrm{max}}+1) \ ,
	\end{equation}
	which is smaller than $n_{\mathrm{H}}$ given by eq.~\eqref{eq:n_H_orthogonal}.
	This provides an example of how one may exploit the symmetries embedded in the chosen BLFQ to achieve gains in computational efficiency.
	
	\section{Computing observables from the valence LFWF\label{sec:observ}}
	One of the many advantages of the light-front approach to quantum field theories is that observables for bound states can be easily extracted from light-front wave functions. Explicitly, measurement operators corresponding to physical observables usually take a simple form, resulting in efficient measurements on a quantum computer (see Sec.~\ref{ss:qc_methods} and App.~\ref{app:num_observ}).
	In this section, we demonstrate how to calculate the decay constant, mass radius, valence parton distribution function, and elastic form factor.
	
	\subsection{The decay constant} The meson decay constants are defined as the matrix elements of current operators between the vacuum and the meson wavefunctions~\cite{Li:2017mlw}. They correspond to amplitudes of the wavefunctions at the coordinate-space origion. Specifically, the decay constants for scalar mesons ($f_{\mathrm{S}}$), pseudoscalar mesons  ($f_{\mathrm{P}}$), vector mesons  ($f_{\mathrm{V}}$), and axial vector mesons ($f_{\mathrm{A}}$) are defined as
	\begin{subequations}
		\label{eq:decay_constants}
		\begin{align}
		& \langle 0 \vert \overline{\psi}\,\gamma^\mu\,\psi \vert \mathrm{S}(p)\rangle =p^\mu f_{\mathrm{S}} \ ,\\
		& \langle 0 \vert \overline{\psi}\,\gamma^\mu\gamma_5\,\psi \vert \mathrm{P}(p)\rangle =i\,p^\mu f_{\mathrm{P}} \ ,\\
		& \langle 0 \vert \overline{\psi}\,\gamma^\mu\,\psi \vert \mathrm{V}(p)\rangle =\epsilon^{\mu}_{\lambda}(p)\,m_{\mathrm{V}} f_{\mathrm{V}} \ ,\\
		& \langle 0 \vert \overline{\psi}\,\gamma^\mu\gamma_5\,\psi \vert \mathrm{A}(p)\rangle =\epsilon^{\mu}_{\lambda}(p)\,m_{\mathrm{A}} f_{\mathrm{A}} \ ,
		\end{align}
	\end{subequations}
	respectively. Here the polarization vector for the vector mesons is defined as
	\begin{equation}
	\epsilon^{\mu}_{\lambda}(p)=
	\begin{cases}
	\left(\frac{p^+}{m_{\mathrm{V},\mathrm{A}}},\,\frac{\vv{p}^{\perp 2}-m_{\mathrm{V},\mathrm{A}}^2}{m_{\mathrm{V},\mathrm{A}}\,p^+},\,\dfrac{\vv{p}^{\perp} }{m_{\mathrm{V},\mathrm{A}}}\right)\quad &\text{for } \lambda=0\\
	\left(0,\,\dfrac{2\vv{e}^{\perp}_\lambda\cdot \vv{p}^\perp }{p^+},\, \vv{e}^{\perp}_{\lambda} \right) \quad & \text{for } \lambda=\pm 1
	\end{cases} \ ,
	\end{equation}
	with $\vv{e}^{\perp}_{\pm}=(1,\pm i)/\sqrt{2}$.
	
	In terms of the valence-sector light-front wavefunctions, expressions for these decay constants are reduced into~\cite{Li:2017mlw}
	\begin{subequations}
		\label{eq:decay_constant_valence_WF}
		\begin{align}
		&\begin{alignedat}{8}
		&f_{\mathrm{P},\mathrm{A}} =2\sqrt{N_{\mathrm{c}}}\int_{0}^{1}\dfrac{dx}{4\pi\sqrt{x(1-x)}}\int \dfrac{d^2 \kappa^\perp}{(2\pi)^2}
		\\&
		\times \left[\psi_{+-}\left(x,\vv{\kappa} ^\perp\right)-\psi_{-+}\left(x,\vv{\kappa} ^\perp\right) \right]\bigg\vert _{m_J=0} \ ,
		\end{alignedat}
		\\
		\label{eq:decay_constant_valence_WF_PA}
		&\begin{alignedat}{8}
		&f_{\mathrm{S},\mathrm{V}} =2\sqrt{N_{\mathrm{c}}}\int_{0}^{1}\dfrac{dx}{4\pi\sqrt{x(1-x)}}\int \dfrac{d^2 \kappa^\perp}{(2\pi)^2}
		\\&
		\times\left[\psi_{+-}\left(x,\vv{\kappa} ^\perp\right)+\psi_{-+}\left(x,\vv{\kappa} ^\perp\right) \right]\bigg\vert _{m_J=0} \ ,
		\end{alignedat}
		\end{align}
	\end{subequations}
	with the condition $m_J=m+s_1+s_2=0$ specifying that only the states with zero angular momentum projections are used in the calculation. Here $N_{c}=3$ is the number of colors.
	
	In the basis representation, the integrals over the longitudinal momentum fraction and the relative transverse momenta in eq.~\eqref{eq:decay_constant_valence_WF} can be evaluated exactly. Details of this calculation can be found in Appendix~\ref{ss:itg_decay_constant}. Since the decay constant is linear in the wave function, we only need to calculate these integrals for each basis function. Subsequently, the decay constants in the basis representation are given by
	\begin{align}
	\label{eq:fpa}
	&\begin{alignedat}{8}
	&f_{\mathrm{P},\mathrm{A}} =2\sqrt{\dfrac{N_{\mathrm{c}}}{\pi}}\,\sum_{n,l}(-1)^n\,L_l(1/2,1/2;\alpha,\beta)
	\\&\times\left( \psi_{n0l+-}-\psi_{n0l-+} \right)\big\rvert _{m_J=0} \ ,
	\end{alignedat}
	\\
	&\begin{alignedat}{8}
	&f_{\mathrm{S},\mathrm{V}} =2\sqrt{\dfrac{N_{\mathrm{c}}}{\pi}}\,\sum_{n,l}(-1)^n\,L_l(1/2,1/2;\alpha,\beta)
	\\&\times\left( \psi_{n0l+-}+\psi_{n0l-+} \right)\big\rvert _{m_J=0} \ ,
	\end{alignedat}
	\end{align}
	where the longitudinal integrals $L_l(a,b;\alpha,\beta)$ are defined and given analytically in Appendix~\ref{ss:def_L_a_b}.
	Because the overall phase of the LFWF remains undetermined by the Hamiltonian, only the absolute value of the decay constant carries physical significance. Once the LFWF $\vert\psi\rangle$ in our basis representation is known on a quantum computer, the calculation of the corresponding decay constant can be thought of as computing $|\langle v|\psi\rangle|$ for some fixed $\vert v \rangle$.

	\subsection{The mass radius}
	The mass radius is the square root of the expectation value for the relative transverse separation of the valence quarks. It can be calculated from the valence two-body wave-function based on eq.~(33) of Ref.~\cite{Li:2017mlw}. Specifically for the pseudoscalar mesons, we have
	\begin{equation}
	\begin{alignedat}{8}
	    \langle r_m^2\rangle =\dfrac{3}{2}\sum_{r,s}&\int_{0}^{1}\dfrac{dx}{4\pi}\int d\vv{r}^\perp\,x(1-x)\vv{r}^{\perp 2}
	    \\&\times
	    \tilde{\psi}^*_{rs}\left(x,\vv{r}^\perp \right)\tilde{\psi}_{rs}\left(x,\vv{r}^\perp\right) \ ,
	\end{alignedat}
	\end{equation}
	where $\tilde{\psi}_{rs}\left(x,\vv{r}^\perp\right)$
	is the light-front wave-function depending on the longitudinal momentum fraction $x$ and the relative transverse coordinate $\vv{r}^\perp$. It is related to the momentum-space wave-function by the Fourier transform in the transverse momenta $\vv{\kappa}^\perp$. Explicitly, we have
	\begin{equation}
	\begin{alignedat}{8}
    \tilde{\psi}_{rs}\left(x,\vv{r}^\perp\right)=&\sqrt{x(1-x)}\sum_{nml}\psi_{nmlrs}\,
    \\&\times
    \tilde{\phi}_{nm}\left(\sqrt{x(1-x)}\vv{r}^\perp \right)\chi_l(x) \ ,	
	\end{alignedat}
	\end{equation}
	with
	\begin{equation}
\begin{multlined}
	\tilde{\phi}_{nm}\left(\vv{r}^\perp \right)
	 =b\sqrt{\dfrac{n!}{(n+|m|)!\pi}}\left(b |\vv{r}^\perp| \right)^{|m|}
\\ \times
	 \exp\left[-\dfrac{b^2\vv{r}^{\perp 2}}{2} \right]
	 L_n^{|m|}\left( b^2\vv{r}^{\perp 2}\right)
\\ \times
	\exp\left[im\phi_r+i\left(n+|m|/2\right)\pi \right] \ ,
\end{multlined}
	\end{equation}
	and $\tan\phi_r=r_2/r_1$.
	
	To calculate the mass radius in terms of expansion coefficients $\psi_{nmls_1s_2}$, we first need to evaluate the following dimensionless integrals of the basis functions:
	\begin{equation}\label{eq:def_I_m}
	\begin{alignedat}{7}
	& I_m(n',m',l';n,m,l) \equiv\int_{0}^{1}dx\,\chi_{l'}(x)\chi_l(x)
	\\& \times
	\int_{0}^{+\infty}d|\vv{r}^\perp|^2\int_{0}^{2\pi}\dfrac{d\phi_r}{8\pi}x^2(1-x)^2\,b^2|\vv{r}^\perp|^2
	\\& \times
	\mathrlap{\tilde{\phi}^*_{n',m'}\left(\sqrt{x(1-x)}\vv{r}^\perp\right)\tilde{\phi}_{n,m}\left(\sqrt{x(1-x)}\vv{r}^\perp\right) \ .}
	\end{alignedat}
	\end{equation}
	We then have the square of the radius given by
	\begin{equation}
\begin{multlined}
	\langle r_m^2\rangle =\dfrac{3}{2b^2}\sum_{rs}\sum_{n'm'l'nml}\psi^*_{n'm'l'rs}
	\\ \times
	I_m(n',m',l';n,m,l)\,\psi_{nmlrs} \ .\label{eq:rm2_mass_radius}
	\end{multlined}
	\end{equation}
	The explicit expression for the matrix $I_m(n',m',l';n,m,l)$ is available in Appendix~\ref{ss:itg_mass_radius}, which takes the form of a hermitian operator in our basis representation.

	\subsection{Parton distribution function of valence quarks}
	The probability of finding a quark inside a meson carrying momentum fraction $x$ is given by
	
	\begin{align}
	\begin{alignedat}{8}
	f(x) =&\dfrac{1}{4\pi\,x(1-x)}
	\\
	&\times \sum_{rs}\int \dfrac{d\vv{\kappa}^\perp}{(2\pi)^2}\,\psi^*_{rs}(x,\vv{\kappa}^\perp)\,\psi_{rs}(x,\vv{\kappa}^\perp)\\
	 =&\dfrac{1}{4\pi}\!\!\!\!\!\sum_{n,m,l\mathrlap{'},\,l,r,s}\!\!\!\!\psi^*_{nml'rs}\,\psi_{nmlrs}\,\chi_{l'}(x)\chi_{l}(x) \ ,
	\end{alignedat}
	\label{eq:valence_PDF_lfwf}
	\end{align}

	which is interpreted as the PDF for the valence quark. The PDF for the valence antiquark is given by ${f(1-x)}$~\cite{Lan:2019vui,Lan:2019rba,Lan:2019img}. We use the solutions of $\psi_{nmlrs}$ defined in eq.~\eqref{eq:psi_rs_basis_expansions} to calculate the valence PDFs of mesons.
	
	Notice that eq.~\eqref{eq:valence_PDF_lfwf} defines a bilinear of the light-front wave-functions in the basis representation. To compute the PDF with the LFWFs obtained from a quantum computer, let us rewrite eq.~\eqref{eq:valence_PDF_lfwf} as
	\begin{subequations}
	\label{eq:pdf_final}
	\begin{gather}
	\label{eq:pdf_rho}
	     f(x) =\dfrac{1}{4\pi}\sum_{l\mathrlap{'},\,l}\rho_{l\mathrlap{'},\,l}\,\chi_{l'}(x)\chi_{l}(x) \ , \\
	     \rho_{l\mathrlap{'},\,l}=\!\!\!\sum_{n,m,r,s}\!\!\psi^*_{nml'rs}\,\psi_{nmlrs} \ .
	\label{eq:rhopdf}
    \end{gather}
	\end{subequations}
	Elements of the density matrix $\rho_{l\mathrlap{'},\,l}$ defined in eq.~\eqref{eq:rhopdf} can be evaluated as the expectation value of the corresponding projection operators on a quantum computer, and subsequently used to calculate the PDF in eq.~\eqref{eq:pdf_rho}.
	
	\subsection{The elastic form factor for pseudoscalar mesons}
	
	To calculate the elastic form factors from the light-front wavefunctions within the impulse approximation where the photon interacts with the meson through the quark-photon vertex, we apply the following formula~\cite{drellyan,West:1970av} within the Drell-Yan frame $P'^{+}=P^+$:
	\begin{equation}\label{eq:def_I_mj,mj'}
    \hspace{-1cm}\begin{alignedat}{8}
	& I_{m_{J}',m_{J}}(Q^2) = \dfrac{1}{{2P^+}}
	\\
	& \mathrlap{ \times \Big\langle \Psi (P',m_J')\Big\vert \sum_{\mathrm{f}}\mathrm{e}_{\mathrm{f}}\,\overline{\psi}_{\mathrm{f}}(0)\gamma^+ \psi_{\mathrm{f}}(0) \Big\vert \Psi(P,m_J) \Big\rangle}
	\\
	& =\sum_{rs}\int \dfrac{dx}{4\pi x(1-x)}\int \dfrac{d^2 k^\perp}{(2\pi)^2}
	\\ &
	\times\bigg\{\mathrm{e}_{\mathrm{q}}\,\psi^{*\,m_{J'}}_{rs}\left(x,\vv{k}^\perp+(1-x)\vv{q}^\perp\right)
	\\ &
	\ \mathrlap{-\mathrm{e}_{\overline{\mathrm{q}}}\, \psi^{*\,m_{J'}}_{rs}\left(x,\vv{k}^\perp-x\vv{q}^\perp\right)\bigg\} \psi^{m_{J}}_{rs}\left(x,\vv{k}^\perp\right) \ ,}
	\end{alignedat}
	\end{equation}
	with $q=P'-P$ and $Q^2=-q^2$. The operator inside the Dirac bracket is the charge density operator on the light front, with $\mathrm{e}_\mathrm{f}$ being the charge carried by the quark of flavor $\mathrm{f}$ in units of the elementary charge and the summation running over all quark flavors. Additionally, $\mathrm{e}_{\mathrm{q}}$ is the charge of the quark ($\mathrm{e}_{\mathrm{u}}=+2/3$ for an up quark). While $\mathrm{e}_{\overline{\mathrm{q}}}$ is the charge of the antiquark ($\mathrm{e}_{\overline{\mathrm{d}}}=-1/3$ for an anti-down quark). Detailed derivation of eq.~\eqref{eq:def_I_mj,mj'} is given in Appendix~\ref{eq:red_Drell_Yan_West}.
	
	In the basis representation, we apply the Talmi-Moshinsky transform to simplify the integrals in the transverse momentum, leaving the longitudinal integral to be evaluated numerically for each $Q^2$. Following steps in Appendix~\ref{ss:Drell_Yan_West_basis_rep}, we rewrite the electromagnetic form factors into a bilinear form of the valence wave-function:
	\begin{equation}
	\begin{alignedat}{8}
	I_{m_J',m_J}(&Q^2) = \sum_{n'm'l'} \sum_{n,m,l} \sum_{r,s}\,\psi^*_{n'm'l'rs}
	\\&
	\mathrlap{ \times \tilde{C}(n',m',l';n,m,l;Q^2)\,\psi_{nmlrs} \ .}
	\end{alignedat}
	\end{equation}
	The operator $\tilde{C}$ is defined according to eq.~\eqref{eq:def_c_tilde_DYW_basis_rep}. At a given $Q^2$, the form factor can be calculated using the LFWFs obtained from a quantum computer by taking the expectation value of the Hermitian operator $\tilde{C}(n',m',l';n,m,l;Q^2)$.
	
	Specifically, the elastic form factors of the pseudoscalar mesons are given by
	\begin{equation}
	\label{eq:def_F_pseudoscalar}
	F_{\mathrm{P}}(Q^2)=I_{0,0}(Q^2) \ .
	\end{equation}
	The charge radius is then specified by the first Taylor expansion coefficient of the elastic form factor at the origin:
	\begin{equation}
	\label{eq:def_charge_radius_pseudoscalar}
	\langle r_{\mathrm{c}}^2 \rangle= -6\lim\limits_{Q^2\rightarrow 0}\,\dfrac{d}{dQ^2}\,F_{\mathrm{P}}(Q^2) \ .
	\end{equation} 

\section{Quantum-Computational Methods\label{ss:qc_methods}}

\begin{figure}[htp]
\centering
\begin{tikzpicture}[node distance=1cm, auto,]
    \node[name=dummy] (dummy) {};
    \node[punkt, name=classical, left = 1.1cm of dummy] (classical) {Classical Minimization Algorithm};
    \node[punkt, right = 1.1 of dummy] (quantum) {Quantum Computer};
    \draw [-{Latex[length=3mm,width=2mm]}] (classical) edge[bend left=15] ["\makecell{Updated\\parameters $\vv{\theta}$}"]  (quantum);
    \draw [-{Latex[length=3mm,width=2mm]}] (quantum) edge[bend left=15] ["\makecell{The value of the\\cost function\\$E(\vv{\theta})=\sand{\psi(\vv{\theta})}{H}{\psi(\vv{\theta})}
    $}"]  (classical);
\end{tikzpicture}
\caption{Schematic of the Variational Quantum Eigensolver (VQE). The parameter vector $\vv{\theta}$ completely specifies the ansatz wave function.}
\label{fig:VQE}
\end{figure}
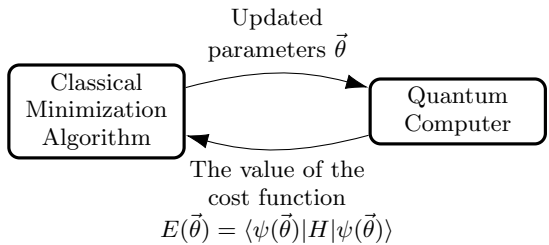

The standard approach to quantum simulation of Hamiltonian dynamics of QFTs is as follows~\cite{2011arXiv1112.4833J}: a) initialize the system in a certain state of the free Hamiltonian, b) adiabatically turn on the interaction, c) if necessary, evolve the system with time, and d) measure the energy (or another observable) of the system using the phase estimation algorithm~\cite{lloyd1996universal,aspuru2005simulated,2011arXiv1112.4833J}.
However, near-term devices cannot perform this procedure due to
limits in qubit numbers and in gate fidelities.
This motivates the variational quantum eigensolver (VQE), an approach to finding Hamiltonian eigenvalues in which a NISQ device is used as a part of a hybrid quantum-classical algorithm~\cite{peruzzo2014variational}.
In VQE, a quantum comp  uter prepares a given variational state and evaluates the Hamiltonian expectation value, which a classical computer performs a gradient search to minimize (see Fig.~\ref{fig:VQE}).
To prepare the variational state, we adopt an ansatz specified by parameters $\vv{\theta}$, which are controlled by the classical minimization.

While in~\cite{Kreshchuk:2020dla} we focused on \emph{ab initio} simulations which are likely to become available in the fault-tolerant regime, in this paper we investigate the use of NISQ devices for high-energy nuclear physics calculations on the light front.
Therefore, unlike in~\cite{Kreshchuk:2020dla}, we formulate the problem as a VQE instance.

We begin by briefly reviewing the VQE method.
For the VQE algorithm to be efficient and accurate, it is essential to come up with a parametrized ansatz state $\sket{\psi(\vv{\theta})}$ that is easy to prepare and is expected to have significant overlap with the true ground state.
Below we shall consider different choices of ansatz state preparation prodecures, encodings of the physical states in a quantum computer, and classical optimization algorithms.

While using VQE for simulating a Hamiltonian problem, the major steps are:
\begin{enumerate}
    \item Define the state/operator mapping, \emph{i.e.}, a correspondence between the physical states and the multi-qubit states of a quantum computer, as well as the mapping between the operators acting on these spaces.

    \item Choose a parametrized ansatz state. One typically writes the ansatz state as
    \begin{equation}
        \sket{\psi(\vv{\theta})} = U(\vv{\theta}) \sket{\psi_0} \ ,
    \end{equation}
    where $\sket{\psi_0}$ is a fixed reference state, and $U(\vv{\theta})$ is the VQE ansatz operator.
    One possibility is to choose the form of $U(\vv{\theta})$ to resemble the form of the Hamiltonian evolution operator~$\me^{\iu Ht}$~\cite{Romero_2018}.

    \item

    Once the state $\sket{\psi(\vv{\theta})}$ is prepared for a given set of parameters $\vv{\theta}$, one evaluates the cost function by measuring the expectation value of the multi-qubit Hamiltonian operator:
    \begin{equation}
\label{eq:functomin}
E(\vv{\theta}) =
\sand{\psi(\vv{\theta})}{\widehat{H}}{\psi(\vv{\theta})}
\ .
\end{equation}
    The algorithm can only be considered efficient if the number of measurements grows polynomially with the problem complexity (discussed further below).

    \item The value of the cost function is then sent to the classical optimizer, which either determines the set of parameters for the next iteration of the algorithm, or terminates the algorithm if the desired precision has been achieved.

\end{enumerate}

We shall explore two approaches to simulating problems in the BLFQ formulation, based on two different encoding schemes.
The first of these is the \emph{direct encoding}, widely used in quantum chemistry~\cite{aspuru2005simulated,somma2002simulating}. In such an encoding scheme, one assigns a particular set of qubit registers to each physical (basis) degree of freedom. In application to purely fermionic systems, one may use one qubit to encode one fermionic second-quantized mode, which leads one to the Jordan-Wigner (JW) encoding~\cite{jordanwigner}. Thus, one needs $N$ qubits in order to encode $N$ fermionic modes. The fermionic raising and lowering operators are represented by \mbox{$N$-local} multi-qubit operators, due to the need to enforce anticommutation relations. One can alternatively employ the Bravyi-Kitaev (BK) encoding~\cite{bravyi2002fermionic,32JCP,BK2015,whitsuper} that uses $N$ qubits to store $N$ fermionic modes, with operators being only \mbox{$\log N$-local}.
Circuits implementing VQE ansatz operators are typically  based on trotterization~\cite{Romero_2018}.

The second encoding we employ is \emph{compact encoding}, as was explored in~\cite{Kreshchuk:2020dla} for front-form physics, and in the context of quantum chemistry in~\cite{aspuru2005simulated}. The idea is to only store the occupied modes of multi-particle Fock states.
With this encoding, one can simulate time evolution using sparsity-based techniques, which are optimal in all parameters~\cite{Kreshchuk:2020dla,babbush2016exponentially, Toloui,berry2015hamiltonian,low17a,qubitization}.
These methods, however, require too many gates to be used to prepare ans\"atze on NISQ devices.
Instead, we can use arbitrary state preparation as long as we restrict to small fixed numbers of particles in the system, as discussed below.

\subsection{Direct encoding\label{sec:direct}}

In order to run a simulation on a current quantum device, in this work we consider a scenario where the particle number is fixed. However, since quantum advantage is likely to be achieved only in the multi-particle regime, it is essential for our methods to be extendable to this more general scenario.

A natural way of formulating a multi-particle problem is by using the second-quantized formalism.
Consider a Hamiltonian of the form
\begin{equation}
    \label{eq:HH}
    \widehat{H} = \widehat{H}_1 + \widehat{H}_2 + \ldots \ ,
\end{equation}
where
\begin{equation}
    \label{eq:h12}
    \widehat{H}_1 = \sum_{i,j} h_{ij} \OPad_i \OPa_j  \ , \;\;\,
    \widehat{H}_2 = \sum_{i,j,k,l} h_{ijkl} \OPad_i \OPad_j \OPa_k \OPa_l \ .
\end{equation}
Here $h_{ij}$ represents the single-body interactions, while $h_{ijkl}$ and higher-order terms correspond to many-body interactions. For the first experimental implementation, we restrict ourselves to ${\widehat{H}=\widehat{H}_1}$ (with $h_{ij}$ being the meson valence sector BLFQ Hamiltonian matrix), for two reasons. First, owing to the efficiency of the BLFQ formulation, considering the single-body part of the Hamiltonian is oftentimes enough to give reasonably good results~\cite{positronium,LI2016118,basislightmesons,Lan:2019vui,Lan:2019rba}. Second, this is suitable for benchmarking, paralleling state-of-the-art experimental results in quantum simulation of chemistry~\cite{google20a}.

Within the JW encoding, the multi-qubit states $\sket{\ldots f_2f_1f_0}$ mimic the second-quantized fermionic states: the qubit $f_i$ stores the occupancy of the $(i+1)$-th orbital (see Table~\ref{tab:encoding}).
In order to enforce anticommutation relations, the fermionic creation and annihilation operators are represented by $N$-local multi-qubit operators~\cite{jordanwigner}. We shall use this encoding for the rest of the section; simulation in the Bravyi-Kitaev encoding is discussed in App.~\ref{app:BK}.

\begin{table}[htp]
    \centering
    \begin{tabular}{|l|l|l|}
    \hline
         Basis state index & Direct encoding & Compact encoding \\ \hline
         $1$ & $\sket{0001}$ & $\sket{00}$ \\ \hline
         $2$ & $\sket{0010}$ & $\sket{01}$ \\ \hline
         $3$ & $\sket{0100}$ & $\sket{10}$ \\ \hline
         $4$ & $\sket{1000}$ & $\sket{11}$ \\ \hline
    \end{tabular}
    \caption{The multi-qubit representation of the four physical one-particle states in the direct and compact encodings.}
    \label{tab:encoding}
\end{table}

Since one typically solves the problem in a basis found by means of some classical approximation, the reference state $\sket{\psi_0}$ can be chosen to have a simple form in terms of basis vectors; in the simplest case, it may coincide with one of the basis vectors. Next, we would like to design an ansatz operator that acts on the reference state to prepare an ansatz state that ideally has large overlap with the exact ground state. An example of such an operator is the Unitary Coupled Cluster (UCC)~\cite{Romero_2018}. Choosing the form of the ansatz operator to resemble the form of the Hamiltonian ensures that one can explore the regions of the Hilbert space that can be reached via the Hamiltonian time evolution, and also guarantees that the symmetries are preserved.
For the Hamiltonian of the form~\eqref{eq:h12} one writes the UCC as~\cite{Romero_2018}
\begin{equation}\label{UCCS}
\begin{gathered}
U(\vv{\theta}) = \me^{T-T^\dagger} \ , \;\;\,
T = T_1 + T_2 + \ldots \ ,
\\
T_1 = \!\!\!\sum_{\substack{i\in\rm{occ}\\a\in\rm{virt}}} \! \theta^{i}_{a} \OPad_a \OPa_i \ , \;\;\,
    T_2 = \!\!\!\!\!\sum_{\substack{i>j\in\rm{occ}\\a>b\in\rm{virt}}} \!\!\!\! \theta^{ij}_{ab} \OPad_a \OPad_b \OPa_i \OPa_j \ ,
    \end{gathered}
\end{equation}
where $\rm{occ}$ and $\rm{virt}$ denote occupied and unoccupied orbitals in the reference state~$\sket{\psi_0}$. Physically, the action of the UCC operator allows one to transfer ``some amplitude'' from initially occupied orbitals to the unoccupied ones. For real Hermitian Hamiltonians, the coefficients in~\eqref{UCCS} are real.

We would now like to translate~\eqref{UCCS}, which was written in terms of the fermionic operators, into its qubit representation.
According to the JW transformation~\cite{jordanwigner}, the qubit operators are obtained as
\begin{equation}
\label{eq:JWmapterm}
\begin{alignedat}{8}
\OPad_j& \OPa_1 - \OPad_1 \OPa_j
    \\
    &\mapsto
    \dfrac{\iu}{2} Y_1 Z_2 \ldots Z_{j-1} X_j -
    \dfrac{\iu}{2} X_1 Z_2 \ldots Z_{j-1} Y_j
    \ ,
\end{alignedat}
\end{equation}
where $X_i,Y_i,Z_i$ are the Pauli matrices acting on qubit $i$.
Substituting \eqref{eq:JWmapterm} into \eqref{UCCS} generates a mapping:
\begin{equation}
    \label{eq:UCCevol}
    U(\vv{\theta}) \mapsto
    \me^{ \iu \sum_j \alpha_j P_j } \ ,
\end{equation}
where $P_j$ are the Pauli operators, while $\alpha_j$ are the corresponding real coefficients. Trotterization of the expression above leads to
\begin{equation}
    U(\vv{\theta}) \mapsto
    \left(\prod_j
    \me^{ \iu \frac{\alpha_j}{\rho} P_j }\right)^\rho \ ,
\end{equation}
where $\rho$ is the Trotter number, which can be typically chosen quite small in VQE~\cite{wecker2015progress}, unlike the case of simulating time evolution.

The traditional approach to calculating the expectation value as in~\eqref{eq:functomin} amounts to expanding the Hamiltonian in the basis of Pauli operators using~\eqref{eq:JWmapterm}:
\begin{equation}
    \label{eq:HPauliexp}
    \sand{\psi(\vv{\theta})}{\widehat{H}}{\psi(\vv{\theta})}
    = \sum_i h_i \sand{\psi(\vv{\theta})}{P_i}{\psi(\vv{\theta})} \ .
\end{equation}
The expectation values of individual Pauli terms on the RHS of~\eqref{eq:HPauliexp} can be
efficiently measured via sampling from the state $\sket{\psi(\vv{\theta})}$~\cite{peruzzo2014variational}.

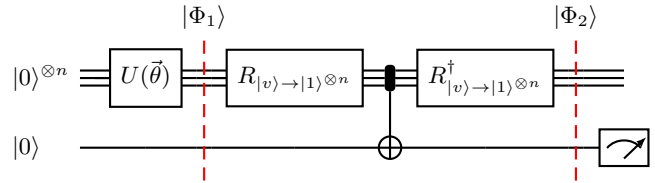
\begin{figure}[htp]
\begin{tikzpicture}
\hspace{-.6cm}
\node[scale=1]{
\begin{quantikz}[slice label style ={inner sep=10pt,anchor=south west,rotate=40}, column sep = .2cm, row sep = .2cm]
\lstick{$\sket{0}^{\otimes n}$} &\qwbundle[alternate]{} & \gate{U(\vv{\theta})
\vphantom{R_{\sket{v}\to\sket{1}^{\otimes n}}^\dagger}
} \qwbundle[alternate]{}
& \qwbundle[alternate]{}
\slice{$\sket{\Phi_1}$}
&\qwbundle[alternate]{} &
\gate{R_{\sket{v}\to\sket{1}^{\otimes n}}
\vphantom{R_{\sket{v}\to\sket{1}^{\otimes n}}^\dagger}
}
\qwbundle[alternate]{}
&  \ctrlbundle{1}&
\gate{R^\dagger_{\sket{v}\to\sket{1}^{\otimes n}}}
\qwbundle[alternate]{}
&\qwbundle[alternate]{}\slice{$\sket{\Phi_2}$}&\qwbundle[alternate]{}&\qwbundle[alternate]{} \\
\lstick{$\sket{0} \hphantom{{}^{\otimes n}}$}& \qw & \qw & \qw & \qw & \qw & \targ{} & \qw &\qw&\qw & \meter{}
\end{quantikz}
};
\end{tikzpicture}
\caption{
Estimating the magnitude of the inner product $\sbraket{v}{\psi(\vv{\theta})}$ for fixed $|v\rangle$.
Up to the first dashed line, the circuit prepares the VQE ansatz state by applying the ansatz circuit $U(\vec{\theta})$ to the first set of registers, resulting in the state ${\sket{\Phi_1} = \sket{\psi(\vv{\theta})}\otimes\sket{0}}$.
The next rotation,
$R_{\sket{v}\to\sket{1}^{\otimes n}}$,
represents any unitary operator that maps the state $|v\rangle$ to the state $|1\rangle^{\otimes n}$.
Thus the state $\sket{\Phi_2}$ at the second dashed line is given by ${\sket{\Phi_2} = \sbraket{v}{\psi(\vv{\theta})}\sket{v}\otimes\sket{1}+(\sket{\psi(\vv{\theta})}-\sbraket{v}{\psi(\vv{\theta})}\sket{v})\otimes\sket{0} }$.
The quantity ${\bigl|{\sbraket{v}{\psi{(\vv{\theta})}}\bigr|}=\sqrt{\bigl|{\sbraket{v}{\psi(\vv{\theta})}\bigr|}^2}}$ is found as the square root of the probability for the \emph{ancilla} qubit to collapse into the state $\sket{1}$.
}
\label{fig:linobservmeas}
\end{figure}

The optimal parameters $\vv{\theta}{}^*$, obtained upon successful termination of the VQE algorithm, allow one to prepare the VQE approximation to the ground state of the system. By analogy with~\eqref{eq:HPauliexp}, this can be used to calculate the expectation value of any observable bilinear in the wave function (\emph{i.e.},~of the form ${\sand{\psi(\vv{\theta})}{\widehat{O}}{\psi(\vv{\theta})}}$), such as mass radius, PDF, or elastic form factor. Observables linear in the wave function (\emph{i.e.},~of the form ${\bigl|\sbraket{v}{\psi(\vv{\theta})}\bigr|}$, where $\sket{v}$ is a constant vector), such as the decay constant, eq.~\eqref{eq:decay_constant_valence_WF_PA}, can be calculated using the simple circuit shown in Fig.~\ref{fig:linobservmeas}.

\subsubsection*{Efficiency analysis}

When proposing new algorithms for quantum simulations on NISQ devices, it is essential to elicit their scaling properties in order to distinguish aspects of a particular simulation that may lead to quantum advantage from those that cannot. The question relevant to the present paper is which aspects of the few-qubit calculations we can scale up to several hundred qubits. Concomitantly, which aspects of the few-qubit calculations are amenable to efficient classical calculations and which are not.

The stages of one shot of a VQE calculation are ansatz preparation and measurement of all Hamiltonian terms. Many shots with fixed ansatz parameters are required to obtain one estimate of the expectation value of the Hamiltonian. Many estimations of the expectation value of the Hamiltonian are required to optimize the ansatz parameters. Typically the resources required to optimize a given VQE ansatz to a fixed precision cannot be bounded theoretically. This is what makes VQE a heuristic method. However we can determine the computational cost of each step in a single shot and ensure that the quantum gates and qubits required scale polynomially with the problem size. We can also be sure that no known efficient classical algorithm exists for large-scale versions of the problem.

As a prototypical example, consider the problem of finding a ground state in quantum chemistry using VQE. The parameters describing the complexity of the problem are the total number of orbitals, $N$, and the number of electrons in the system, $M$ (\emph{i.e.}, the number of occupied orbitals). Using the direct mapping requires $N$ qubits for encoding physical states. The second-quantized Hamiltonian operator can be written as a polynomial in ladder operators. Those, in turn, are each represented by a polynomial number of Pauli operators, each of which is at most \mbox{$\log N$-local}. Therefore, the measurement of the Hamiltonian operator can be replaced with the measurement of a polynomial number of elementary operators. Each of those can be measured with precision $\epsilon$ using $O(\epsilon^{-2})$ samples~\cite{peruzzo2014variational,mcclean2016theory}.

All that remains is to quantify the operational resources for preparing the ans\"atze.
In quantum chemistry in the direct mapping, these are typically prepared using a unitary coupled cluster (UCC) operator.
The UCC ansatz operator including single and double excitations (UCCSD) contains $O(N^2M^2)$ free parameters. Application of the (trotterized) ansatz operator to the initial state is realized by a circuit containing a polynomial number of gates, since the action of fermionic ladder operators can be represented by a polynomial number of gates in the case of direct encoding.

Let us now see how these arguments can be naturally extended to the case of quantum field theory (QFT). First of all, we note that unlike in quantum chemistry, where the number of particles is conserved, QFT allows for processes of creation and annihilation of particles. Nevertheless, QFT does have an operator similar to the non-relativistic number operator~--- namely, the total momentum operator. Indeed, since this relativistic momentum operator commutes with the Hamiltonian, one can solve the problem within a Fock space sector of a fixed total momentum.

This analogy extends to the terms in the Hamiltonian.
In quantum chemistry, the Hamiltonian operator can be written as a polynomial of fermionic creation and annihilation operators, containing $O(\poly (N))$ terms.
In QFT, the second-quantized Hamiltonian operator can be written as a polynomial of ladder operators, containing $O(\poly (\Lambda))$ terms, where $\Lambda$ is the momentum cutoff.
As in chemistry, in the direct encoding those can be represented by $\mbox{$\log \Lambda$-local}$ Pauli operators, whose total number consequently also scales as $O(\poly (\Lambda))$.
To obtain a finite-dimensional Hilbert space in the equal time quantization, one would have to impose an additional cutoff on the number of excitations in each bosonic mode. However, in the LF formalism the maximum number of excitations is automatically limited by \emph{harmonic resolution} $K$, the dimensionless light-cone momentum~\cite{BRODSKY1998299,Kreshchuk:2020dla}. Within the BLFQ, the role of $\Lambda$ and $K$ is played by $N_{\text{max}}$, $M_{\text{max}}$, and $L_{\text{max}}$ cutoffs (introduced in Sec.~\ref{ss:basis_exp}). Therefore, all the resources for a single VQE estimation of the QFT Hamiltonian expectation value based on the direct encoding and UCC will grow polynomially in momentum cutoffs and precision.

\subsection{Compact encoding\label{sec:compact}}

In our previous work~\cite{Kreshchuk:2020dla} we explored the possibility of using the compact encoding for simulating physics on the light front. This amounts to only storing information about occupied modes in the Fock states. In the simplified setting considered in Sec.~\ref{sec:blfq-njl}, due to the usage of relative coordinates, the only information we store is the index of the single occupied orbital. While in the direct mapping the index of the occupied orbital was stored in the unary form, requiring $N$ qubits for $N$ orbitals,
in the compact mapping it is stored in the binary form, requiring $\lceil\log_2 N\rceil$ qubits for $N$ orbitals.
Therefore, in the case when the single-body Hamiltonian matrix $h_{ij}$ is of size ${N\times N=2^n\times2^n}$, one would use all the basis states of the $2^n$-dimensional Hilbert space of $n$ qubits (see Table~\ref{tab:encoding}).

In the compact mapping, an equation analogous to~\eqref{eq:JWmapterm} would contain an exponential number of terms on the RHS, thus making the usage of the UCC inefficient. Instead, one may employ any of existing arbitrary state preparation algorithms~\cite{synth}. While their complexity is exponential in the number of qubits, in our case the number of qubits is itself logarithmic in the problem cutoffs.

For the direct encoding, in order to measure the expectation value of the Hamiltonian we can express the Hamiltonian in terms of Pauli operators. Any observable of size ${N\times N=2^n\times 2^n}$ can be expanded in the basis of $4^n = N^2$ Pauli operators defined on $n$ qubits:
\begin{equation}
    \label{eq:trace}
    h = \sum_{\alpha = 1}^{N^2 } c_\alpha P_\alpha \ ,\;\;\, {c_\alpha = \frac{1}{2^{n}}\tr (h P_\alpha)} \ .
\end{equation}
where $c_\alpha$ are \emph{real}.

It should be emphasized that the logarithmic scaling of the number of qubits required as a function of the problem cutoffs implies that the Hilbert space dimension is polynomial in the cutoffs. This also implies that classical approaches to this problem are efficient. We are considering these specific initial problems as benchmarks where the results obtained can be compared to the known classical solution as an evaluation of the NISQ device itself.

\subsubsection*{Efficiency analysis}

Within the VQE regime, the approach to quantum simulation based on the compact mapping is more efficient than the one based on the direct encoding when solving a two-body problem in the  relative variable basis. As one starts to consider the problem in the multi-particle setting, the number of qubits required for storing physical states in the compact encoding is nearly optimal~\cite{Kreshchuk:2020dla}.
Despite that, one faces serious problems at the stages of state preparation and measurement. Since the complexity of arbitrary state preparation algorithms scales exponentially with the number of qubits, and the number of qubits grows linearly with the number of occupied modes, those algorithms can only be used if the number of particles is fixed and small. Of course, in principle, one could use sparsity-based techniques for state preparation, but this produces gate counts that are not feasible in the NISQ era. Therefore, coming up with a good ansatz for a multi-particle state in the compact encoding is an important task, which we leave for future work.

Another problem arises at the measurement stage: the number of Pauli terms in the expansion of the Hamiltonian grows exponentially with the number of qubits. This motivates the development of VQE techniques for sparse Hamiltonians.

\section{Results}

In this section we describe numerical and experimental results of implementing VQE for a sample QFT problem, namely simulation of a pion in the minimal BLFQ representation.
In order to run our simulation on an existing device, we shall use the $4\times 4$ light meson BLFQ Hamiltonian from Sec.~\ref{sec:blfq-njl} corresponding to ${J_z=0}$ sector in Table~\ref{tab:def_Jz_theta} (see also App.~\ref{ss:H_eff_4_by_4}):
\begin{equation}\label{eq:H_num_4_by_4}
\begin{split}
   &h_{ij} =
H^{\text{BLFQ}}
\\
&\!
=
 \begin{pmatrix}
 640323 & 139872 & -139872 & -107450 \\
 139872 & 346707 &  174794 &  139872 \\
 -139872 & 174794 & 346707 & -139872 \\
 -107450 & 139872 & -139872 & 640323
 \end{pmatrix} \ ,
\end{split}
\raisetag{3\baselineskip}
\end{equation}
in units of MeV$^2$.
The two lowest eigenvalues correspond to $\pi$ and $\rho$ meson squared masses:
the ground state is ${(0.34,-0.62,-0.62,0.34)^T}$, with $m_\pi^2 = 139.6^2$ MeV$^2$.

We can analyze the VQE calculation in a few steps: a)~Check that the classical optimizer is working correctly. To eliminate any errors arising due to sampling, we begin with evaluating the Hamiltonian expectation values exactly, using the statevector representation. b)~Determine the number of steps required to reach the desired precision when evaluating the expectation value via sampling from the exact distribution. This gives the lower bound on the number of samples, and models the situation of using a ``perfect quantum computer.'' c)~Evaluate expectation values on the IBM Vigo quantum processor. d)~Use error mitigation techniques to postprocess the results obtained on the quantum computer.
We shall perform these steps using both the direct and compact encodings, evaluating the Hamiltonian eigenstate as well as other observables discussed in Sec.~\ref{sec:observ}.

\begin{table*}[htp]
    \centering
    \begin{tabular}{ccccccc}
	\hline
	$\mathbf{m}$ & $\overline{\mathbf{m}}$ & $\kappa$ & $G_\pi$ & $N_{\mathrm{max}}$ & $M_{\mathrm{max}}$ & $L_{\mathrm{max}}$ \\
	\hline
	$337.01~\mathrm{MeV}$ & $337.01~\mathrm{MeV}$ & $227.00~\mathrm{MeV}$ & $250.785~\mathrm{GeV}^{-2}$ & $0$ & $2$ & $0$ \\
	\hline
    \end{tabular}
    \caption{Model parameters for the BLFQ-NJL model. }
    \label{tab:model_parameters}
\end{table*}

The multi-qubit states representing the four physical basis states are shown in Table~\ref{tab:encoding}. The states in the direct encoding can be thought of as JW-encoded states. Therefore, we use the JW transformation for calculating the corresponding multi-qubit Hamiltonian:
\begin{equation}
\label{eq:qubitHdirect}
\begin{multlined}
   H^{\text{BLFQ}}_{\text{direct}}
  = 987031 IIII + 87397 (IXXI + IYYI )
  \\
  - 53725 ( YZZY+ XZZX) - 320161 (IIIZ + ZIII)
  \\
  - 173353 (IZII + IIZI) + 69936 ( IIYY + IIXX
  \\ + YZYI+ XZXI - IYZY - IXZX
  \\
  - YYII - XXII )  \ ,
\end{multlined}
\end{equation}
where each term is a tensor product of single-qubit Pauli matrices $I,X,Y,Z$. (In what follows, we use this convention when expanding Hermitian matrices in Pauli terms acting on qubits.)

\begin{figure*}[htp]
\centering
\subfloat[]{
\adjustbox{valign=b}{
\begin{tikzpicture}
\node[scale=1]{
\begin{quantikz}[column sep = .2cm, row sep = .1cm]
\lstick[wires=4]{$\sket{0}^{\otimes 4}$} & \qw & \qw & \qw & \qw & \gate{R_y(\theta_2)} & \ctrl{1} & \qw
\\
& \qw & \gate{X} & \ctrl{1} & \gate{X} & \ctrl{-1} & \gate{X} & \qw
\\
& \qw & \qw & \gate{R_y(\theta_1)} & \ctrl{-1} & \ctrl{1} & \gate{X} & \qw
\\
& \qw & \qw & \qw & \qw & \gate{R_y(\theta_3)} & \ctrl{-1} & \qw
\end{quantikz}
};
\end{tikzpicture}
}
\label{fig:circuitdir}
}
\hspace{.4cm}
\subfloat[]{
\adjustbox{ valign = b}{
\begin{tikzpicture}
\node[scale=1]{
\begin{quantikz}[column sep = .2cm, row sep = .1cm]
\lstick[wires=2]{$\sket{0}^{\otimes 2}$} & \gate{R_y(\theta_1)} & \gate{X} & \gate{R_y(\theta_3)} & \qw
\\
& \gate{R_y(\theta_2)} & \ctrl{-1} & \qw  & \qw \\[.55cm]
\end{quantikz}
};
\end{tikzpicture}
}
\label{fig:circuitbin}
}
\caption{Ansatz circuits for preparing an arbitrary superposition of single-particle Fock states with real coefficients. For the direct encoding (a), we use a generalization of a circuit from~\cite{doi:10.1002/qute.201900015} for preparation of $W_N$ states. For the binary encoding (b), we use arbitrary state preparation, with all single qubit rotations replaced by $R_y(\theta)$ gates, where $R_y(\theta)$ denotes a single-qubit rotation through an angle $\theta$ about the $y$-axis.
}
\label{fig:circuit}
\end{figure*}
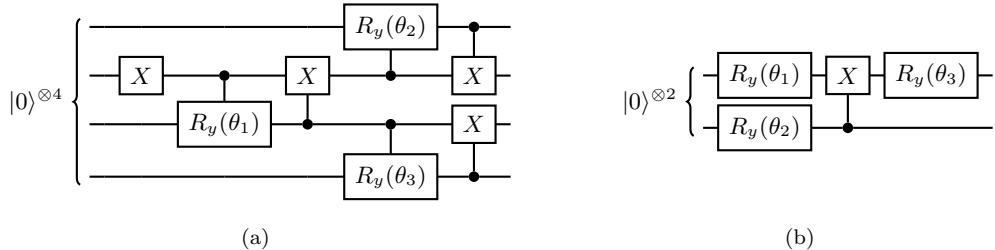

As an ansatz operator, one could use the UCCS (no Doubles) operator. According to \eqref{UCCS}, $T_1$ will contain $N$ terms, each of which is $N$-local in the JW encoding and \mbox{$\log N$}-local in the BK encoding. In the former case, the circuit will contain $O(N^2)$ gates, while in the latter only $O(N\log N)$ gates.
However, in order to further improve the gate count in the direct encoding-based algorithm, instead of the UCCS ansatz, we design a simple parametrized circuit of depth $O(\log N)$ using $O(N)$ gates shown in Fig.~\ref{fig:circuitdir}, which is capable of preparing an arbitrary superposition (with real amplitudes) of single-occupied states in the JW encoding. (This circuit is a generalization of the circuit proposed in~\cite{doi:10.1002/qute.201900015} for preparing $W_N$ states.)

The multi-qubit representation of the Hamiltonian in the compact encoding is obtained from \eqref{eq:H_num_4_by_4}, calculating the coefficients by applying~\eqref{eq:trace}:
\begin{equation}
\label{eq:qubitHcompact}
\begin{multlined}
   H = 33671 XX + 141122 YY + 146807 ZZ\\
     + 493515 II+ 139872 (ZX - XZ ) \ .
\end{multlined}
\end{equation}
The ansatz state is prepared using the circuit shown in Fig.~\ref{fig:circuitbin}, which prepares an arbitrary two-qubit state with real amplitudes.
For both encodings, we need to estimate expectation values of Pauli operators.
To do this, we first rotate to a basis in which the desired Pauli operator is diagonal, then measure single-qubit $Z$ operators.
The desired operator is a product of some set of single-qubit $Z$ operators in this basis.

\begin{sloppypar}
The classical optimization was performed using various algorithms from the \texttt{python.scipy.optimize} library. In agreement with~\cite{Romero_2018}, the best convergence to the true ground state was achieved with L-BFGS-B~\cite{LBFGSB} and COBYLA~\cite{COBYLA} methods. The latter showed better convergence, and it is used in all the following calculations.  Depending on the choice of the initial guess state, the optimizer was typically reaching $4$-digit precision after ${\sim10^1\text{-}10^2}$ steps (for a good initial guess, such as ${(0,-1/\sqrt{2},1/\sqrt{2},0)^T}$) and up to a few hundred steps for randomly chosen initial state. In rare cases, the minimization was not converging.
\end{sloppypar}

\begin{figure}[htp]
\hspace{-.4cm}
\includegraphics[width = .5\textwidth]{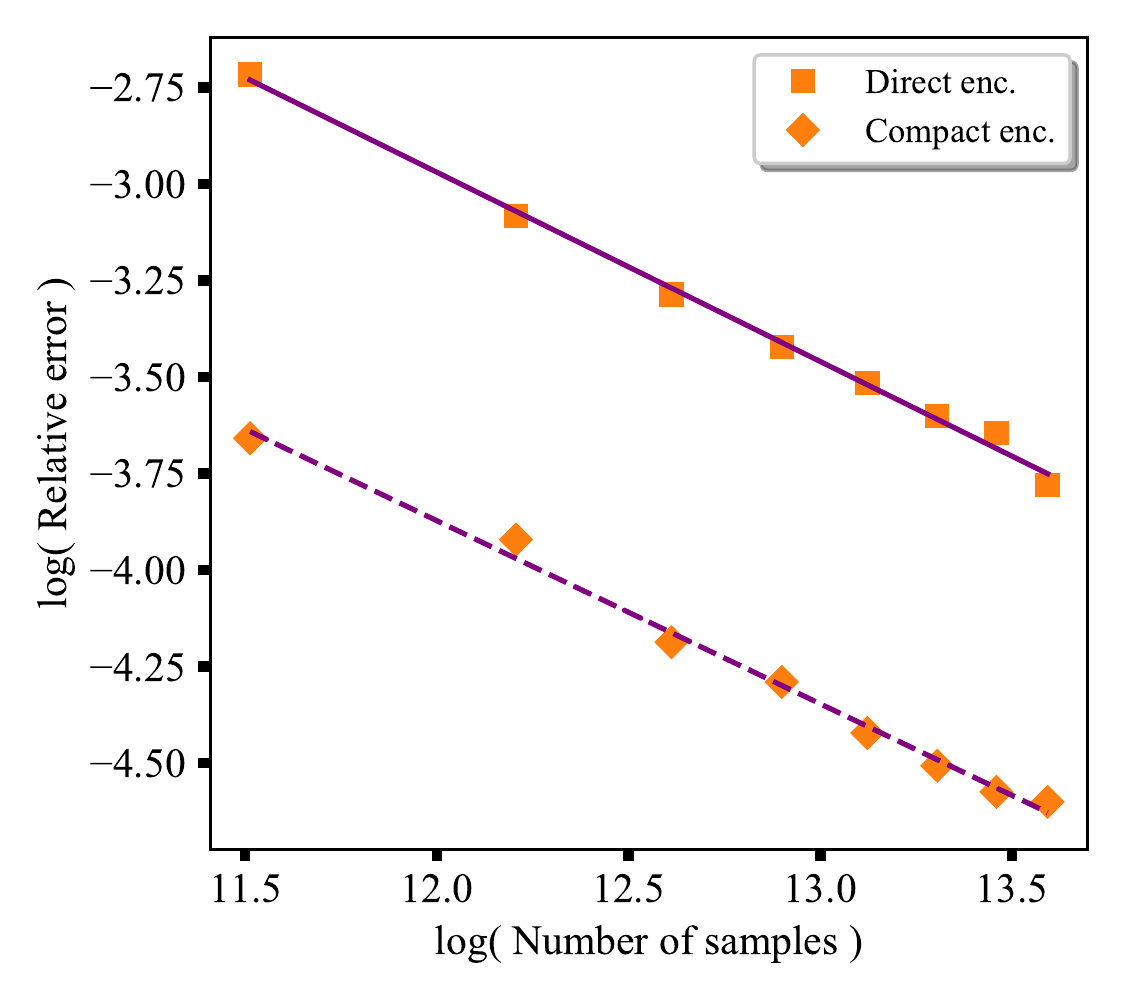}
\caption{Precision vs. number of samples for ground state energy obtained via sampling from the exact distribution. Fitting gives ${n\approx 382/\epsilon^{2.04}}$ (direct encoding) and ${n\approx 46/\epsilon^{2.1}}$ in (compact encoding), confirming the theoretical ${n\sim O(1/\epsilon^2)}$ dependence. Compact encoding shows better convergence due to having shorter circuits on fewer qubits (compare Figs.~\ref{fig:circuitdir} and \ref{fig:circuitbin}).}
\label{fig:errors}
\end{figure}

\begin{figure}[htp]
\hspace{-.4cm}
\includegraphics[width = .5\textwidth]{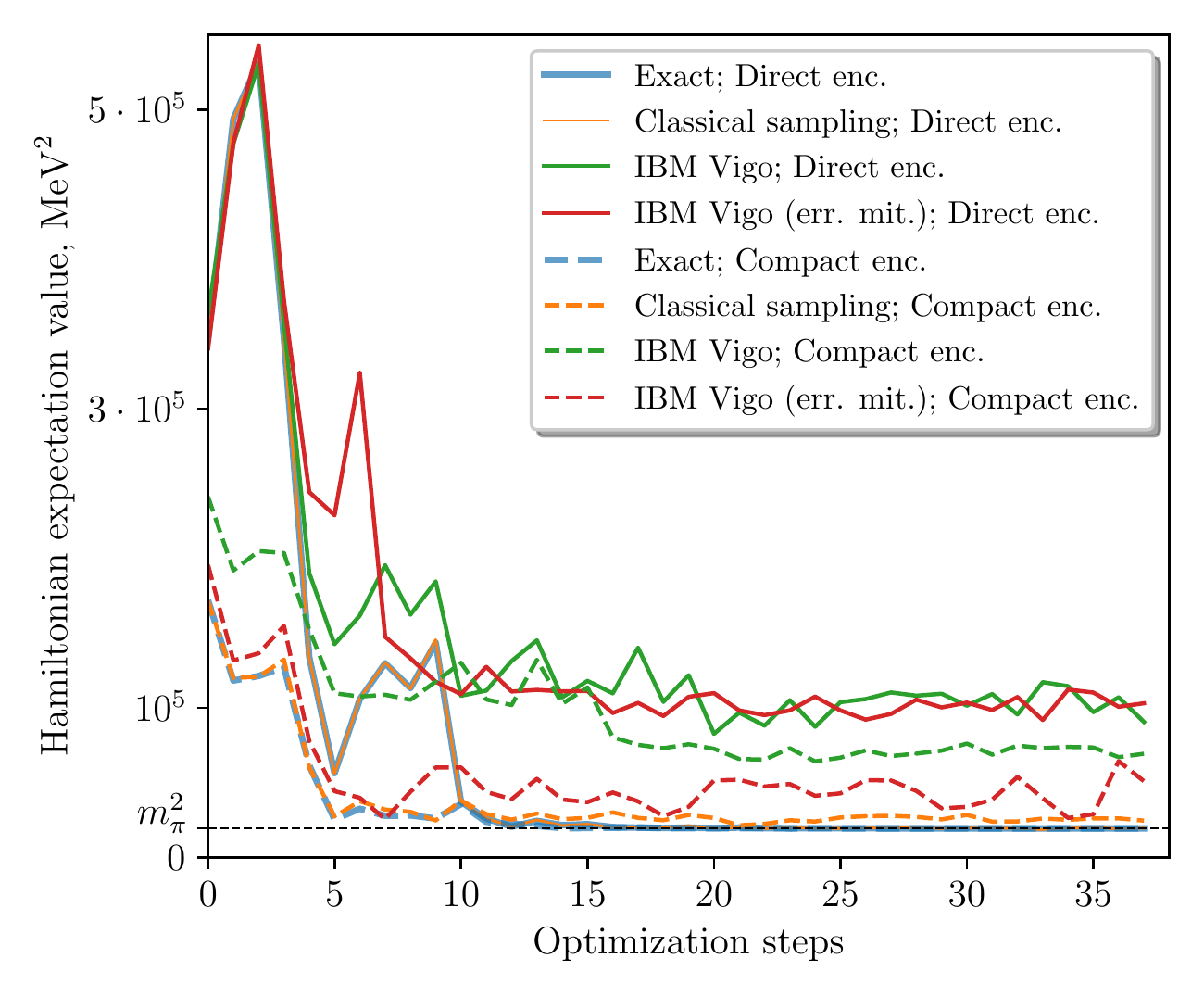}
\caption{The results of the VQE minimization algorithm in the compact and direct encodings. These were obtained from 8192 samples per term on IBM Vigo machine, with and without measurement error mitigation.}
\label{fig:vqe_minimization}
\end{figure}

\begin{figure}[htp]
\hspace{-.4cm}
\includegraphics[width = .5\textwidth]{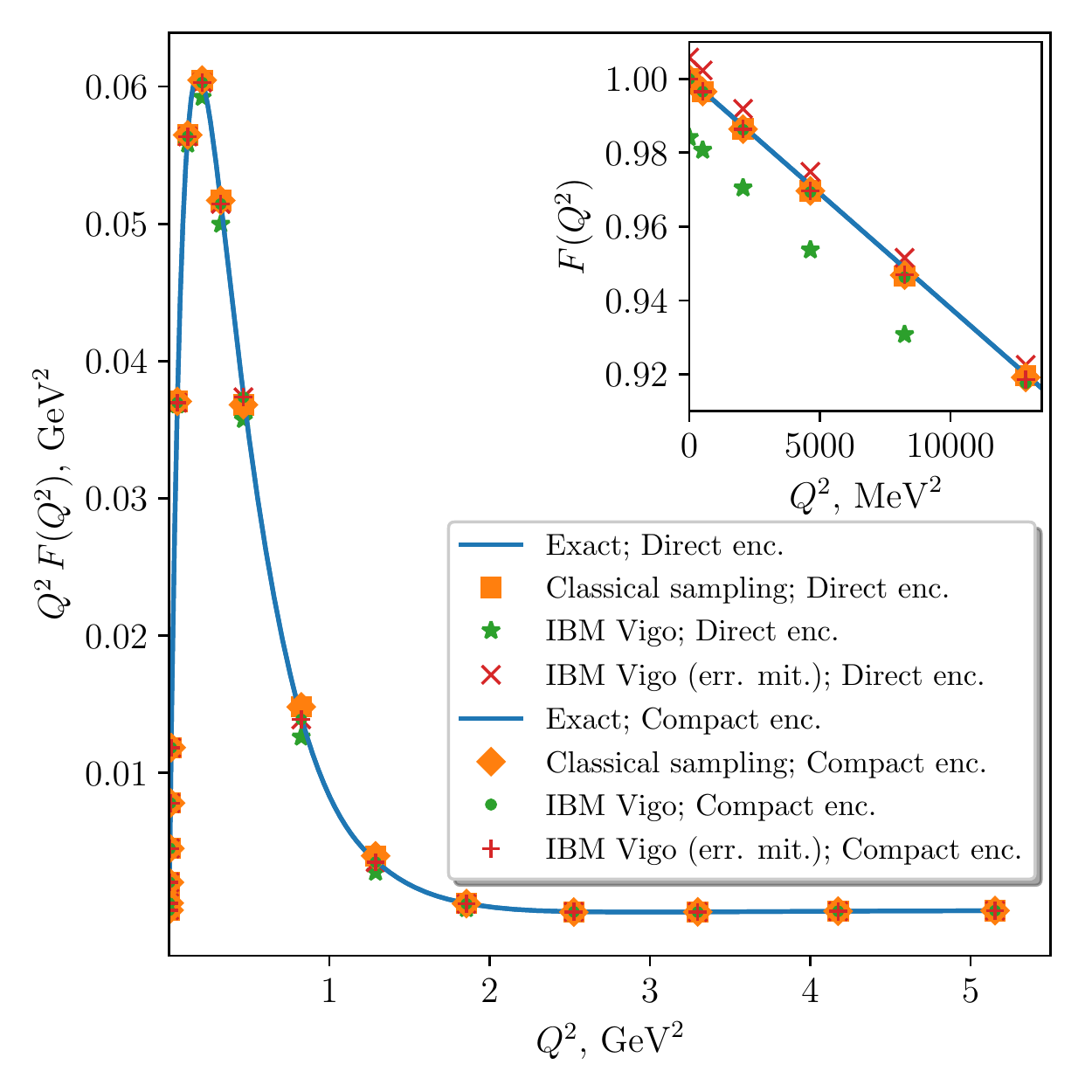}
\caption{Pion elastic form factor, as defined in eq.~\eqref{eq:def_F_pseudoscalar}. Pion elastic form factor is used to calculate the charge radius, obtaining the values given in Tab.~\ref{tab:charge_radius} (charge radius is defined in eq.~\ref{eq:def_charge_radius_pseudoscalar}). Datapoints for the quantum simulation on the IBM Vigo processor used 8192 samples per term, with and without measurement error mitigation. The results measured on the quantum computer are in good agreement with the exact ones due to the strong contribution to the measurement operators from the identity term.}
\label{fig:form_factor}
\end{figure}

Next, we determined the number of samples from the exact distribution required to reach the desired precision, which is expected to scale as $
{O(1/\epsilon^2)}$~\cite{mcclean2016theory}.
To do so, we calculated the relative error for determining the Hamiltonian's expectation value in the true ground state using the classical simulation (the corresponding parameters of the circuits were obtained via the optimization at the previous stage).
We performed $1000$ experiments with a fixed number of samples, and calculated the RMS relative errors in determining the ground state expectation value over each set of experiments.
The results on Fig.~\ref{fig:errors} indicate that on an ideal quantum computer we would need to generate~$\sim10^6$ samples per Pauli term in order to reach $2\%$ precision, and $\sim4\cdot10^6$ samples to reach $1\%$ precision.

\begin{figure*}[htp]
\centering
\includegraphics[width = 1\textwidth]{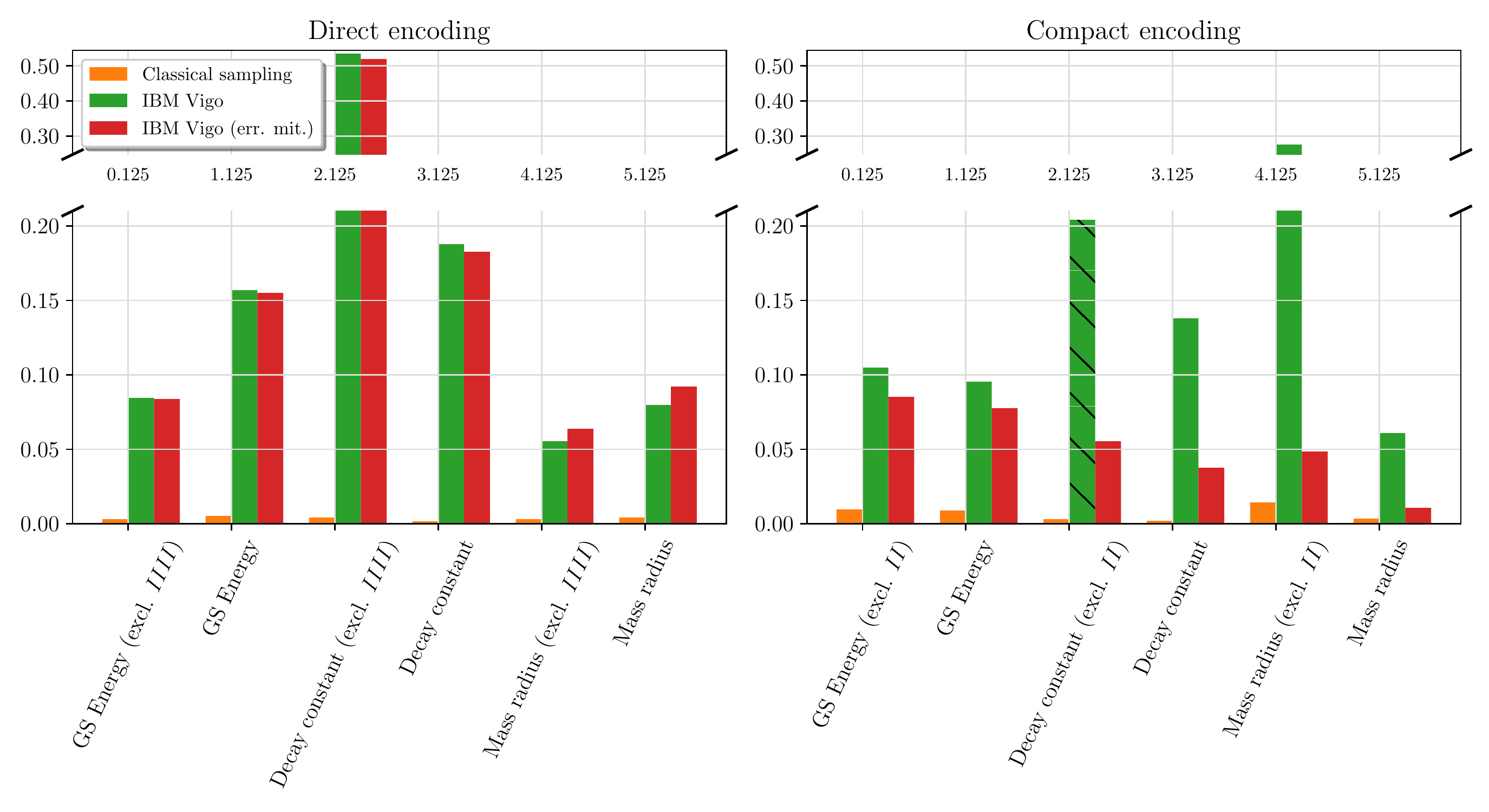}
\caption{Relative errors in estimates of various observables.
These were obtained from 8192 samples per term on IBM Vigo machine, with and without measurement error mitigation.
Physically significant observables have a significant contribution from the constant term in their multi-qubit representation.
Observables are shown with and without the contribution of the constant term.
For the GS energy, the error was calculated relative to the second lowest eigenvalue,~$m^2_\rho$.
For the compact encoding, measurement error mitigation consistently improves the results. }
\label{fig:observ_expects}
\end{figure*}

Fig.~\ref{fig:observ_expects} shows the relative errors for the energy, decay constant, and mass radius, evaluated in the approximate ground state obtained via the VQE minimization procedure.
The expressions for all observables are obtained from the corresponding BLFQ matrices in analogy with eqs.~\eqref{eq:qubitHdirect} and~\eqref{eq:qubitHcompact}; the explicit expressions can be found in App.~\ref{app:num_observ}.
Note that all the observables have a dominant contribution from the unity term ($IIII$ in the direct encoding and $II$ in the compact encoding), whose expectation value is exactly $1$.
Therefore, in Fig.~\ref{fig:observ_expects} we also show the expectation values for observables from which this term has been subtracted, which in certain cases improves the relative precision of results.
The expectation values without the unit terms are the quantities actually measured on the quantum computer, while those including the unit terms are the physically relevant numbers, so the relative errors in both are of interest. In order to calculate the decay constant, one can use the circuit shown in Fig.~\ref{fig:linobservmeas} or Pauli measurements; we use the latter option to minimize the number of gates.

The elastic form factors, eq.~\eqref{eq:def_F_pseudoscalar}, are shown in Fig.~\ref{fig:form_factor}, and the corresponding charge radii, eq.~\eqref{eq:def_charge_radius_pseudoscalar}, are presented in the Table~\ref{tab:charge_radius}. In both cases, the results obtained on the quantum computer are in good agreement with the exact ones. This is to be expected, because the corresponding measurement operators have a large contribution from the identity operator.

With our choice of cutoffs, the calculation of PDFs in the compact encoding reduces to measurement of $II$, while in the direct encoding, it reduces to the projector onto the computational subspace (spanned by the single-occupancy Fock states).
Thus in both cases, the quantity to be measured on the quantum computer is trivial (\emph{i.e.},~${\rho_{0,0}=1}$, as in eq.~\eqref{eq:valence_PDF_lfwf}), and the resulting PDF is ${f(x)=\rho_{0,0}\chi_0^2(x)=1.\times\bigl(2986 \, x^{4.4} (1-x)^{4.4}\bigr)^2}$.

\begin{table}[htp]
\centering
{
\def\arraystretch{1.25}
\begin{tabular}{|l|l|l|l|}
\hline
&  \multicolumn{2}{c|}{Charge radius  $ {\textstyle\sqrt{\langle r_{\mathrm{c}}^2\rangle}}$, MeV$^{-1}$} \\ \hline
Encoding & Direct &  Compact   \\ \hline
Exact & $6.31\cdot10^{-3}$ & $6.31\cdot10^{-3}$ \\ \hline
Classical sampling & $6.29\cdot10^{-3}$ & $6.30\cdot10^{-3}$ \\ \hline
IBM Vigo & $6.33\cdot10^{-3}$ & $6.35\cdot10^{-3}$ \\ \hline
IBM Vigo (err. mit.) & $6.34\cdot10^{-3}$  & $6.31\cdot10^{-3}$ \\ \hline
\end{tabular}
}
\caption{Pion charge radius, as defined in eq.~\eqref{eq:def_charge_radius_pseudoscalar}, calculated using the numerical results from Fig.~\ref{fig:form_factor}.}

\label{tab:charge_radius}
\end{table}

\section{Discussion}

In our paper, for the first time we simulated high energy nuclear physics in the light front formulation on existing devices.
We considered a detailed example in which we studied a relativistic analog of hydrogen, the pion.
We studied this problem in a fixed particle number formulation  as a benchmarking test for existing devices, and as a preparation for moving on to the mixed particle number formulation.
Using the basis light front quantization (BLFQ) formalism, we demonstrated how small quantum computers can be used for calculating baryonic spectra and various observables.
Adopting an effective interaction (suggested by AdS/QCD correspondence~\cite{Brodsky:2008gc}) and a set of basis functions (motivated by the particular problem of interest~\cite{varybasis}) allows one to significantly reduce the computational resources, and to obtain reasonable results for realistic theories on devices having just a few qubits.

Within the VQE approach to quantum simulation, studied in the present work, we considered various encodings and state preparation procedures, some of which were naturally suggested by our experience in quantum chemistry.
Together with our previous paper~\cite{Kreshchuk:2020dla}, this work defines a spectrum of methods for quantum simulation of quantum field theories.
On this spectrum, one can move from restricted models, to be simulated on existing quantum devices, all the way to full \emph{ab initio} simulation of QCD in 3+1D, to be simulated on future fault-tolerant quantum computers.
Future work will expand and improve our methods on both ends of the spectrum.
The next step in near-term simulation will be to switch to single-particle coordinates, providing the framework for mixed particle number simulations where quantum advantage is possible.

\begin{acknowledgements}
W.~M.~K. acknowledges support from the National Science Foundation, Grant No. DGE-1842474.
P.~J.~L.,  M.~K. and G.~G.~acknowledge support from DOE HEP Grant No. DE-SC0019452. S.~J. and J.~P.~V. acknowledge support from DOE Grant Nos. DE-FG02-87ER40371 and DE-SC0018223. S.~J. also acknowledges support from DOE Office of Science, Office of Nuclear Physics, contract No. DE-AC02-06CH11357. This work was supported by the NSF STAQ project (PHY-1818914).
\end{acknowledgements}

\appendix

\section{Hamiltonian in the basis representation}
\subsection{Unitary transformation to the fixed \texorpdfstring{$J_z$}{Jz} blocks\label{ss:ut_fixed_Jz}}
Because the light-front Hamiltonian conserves the angular momentum in the $z$ direction, the Hamiltonian in our basis representation can be diagonalized into blocks of fixed $J_z$. 
This is equivalent to 
combining the spin quantum numbers $s_1$ and $s_2$ with the magnetic quantum number $m$ to form a new quantum number $\theta$. Specifically when $M_{\mathrm{max}} = 2$, the unitary transformation from the original BLFQ basis to this block-diagonal form is given by Table~\ref{tab:def_Jz_theta}. 
\begin{table}[htp]
	\centering
\begin{tabular}{cc|ccc}
	\hline
	$J_z$ & $\theta$ & $m$ & $s_1$ & $s_2$ \\
	\hline
	$-3$ & $1$ & $-2$ & $-$ & $-$ \\
	\hline
	$-2$ & $1$ & $-2$ & $+$ & $-$ \\
	$-2$ & $2$ & $-2$ & $-$ & $+$ \\
	$-2$ & $3$ & $-1$ & $-$ & $-$ \\
	\hline
	$-1$ & $1$ & $-2$ & $+$ & $+$ \\
	$-1$ & $2$ & $-1$ & $+$ & $-$ \\
	$-1$ & $3$ & $-1$ & $-$ & $+$ \\
	$-1$ & $4$ & $0$ & $-$ & $-$ \\
	\hline
	$0$ & $1$ & $-1$ & $+$ & $+$ \\
	$0$ & $2$ & $0$ & $+$ & $-$ \\
	$0$ & $3$ & $0$ & $-$ & $+$ \\
	$0$ & $4$ & $1$ & $-$ & $-$ \\
	\hline
	$1$ & $1$ & $0$ & $+$ & $+$ \\
	$1$ & $2$ & $1$ & $+$ & $-$ \\
	$1$ & $3$ & $1$ & $-$ & $+$ \\
	$1$ & $4$ & $2$ & $-$ & $-$ \\
	\hline
	$2$ & $1$ & $1$ & $+$ & $+$ \\
	$2$ & $2$ & $2$ & $+$ & $-$ \\
	$2$ & $3$ & $2$ & $-$ & $+$ \\
	\hline
	$3$ & $1$ & $2$ & $+$ & $+$ \\
	\hline
\end{tabular}
\caption{The unitary transformation from the orthogonal enumeration to blocks with fixed $J_z$. The other basis quantum numbers $n$ and $l$ are identical in both representations. Here we only impose the basis cut-off of $M_{\mathrm{max}}=2$, while cut-offs in $l$ and $n$ can be freely determined.}
\label{tab:def_Jz_theta}
\end{table}

\subsection{The BLFQ-NJL Hamiltonian in the \texorpdfstring{$J_z=0$}{Jz=0} block\label{ss:H_eff_4_by_4}}
When~$N_{\mathrm{max}}=L_{\mathrm{max}}=0$, the light-front effective Hamiltonian in the $J_z=0$ block takes the form of a $4$-by-$4$ matrix. The subscripts of the matrix then index the basis quantum number~$\theta$. The explicit expressions for these matrix elements are given in the following equations. Here $\kappa$ is strength of the confining potential which we set identical to the basis scale~$b$. The parameter $G_\pi$ is the coupling constant of the NJL interaction. Functions $L'(a,b)$ and $L(a,b)$ both stand for $L_{0}(a,b;\alpha,\beta)$ given in Appendix~\ref{ss:def_L_a_b}.

Explicitly, the matrix elements of this Hamitlonian in our basis representation is given by
\begin{subequations}
\begin{align}
&\begin{alignedat}{7}
H_{11}  = (\mathbf{m}+\overline{\mathbf{m}})^2 &+ 5\kappa^2
\\
&+ \dfrac{8 G_\pi b^4}{\pi}L'(0,0)L(0,0) \ ,
\end{alignedat}
\\
&\begin{alignedat}{7}
H_{12}   = \dfrac{4 G_\pi b^3}{\pi} &\bigg\{ \mathbf{m} \big\{ [L'(0,1)L(1/2,-1/2) \\
& \quad -L'(0,0)L(1/2,-1/2) \big\} \\
& \quad + \overline{\mathbf{m}}\,L'(0,0)L(-1/2,1/2) \bigg\} \ ,
\end{alignedat}
\raisetag{3\baselineskip}
\\
&\begin{alignedat}{7}
H_{13} &  = -\dfrac{2 G_\pi b^3}{\pi} L'(0,0)\bigg\{\overline{\mathbf{m}} \big\{ 2L(-1/2,1/2) \\
&\quad +L(-1/2,3/2)+L(1/2,1/2) \big\}  \\
& \hspace{2.5cm} + 2\mathbf{m}\,L(1/2,-1/2) \bigg\} \ ,
\end{alignedat}
\\
&\begin{alignedat}{7}
H_{14} &  = -\dfrac{4 G_\pi b^4}{\pi}\big\{ L'(0,1)L(1,0) + L'(1,0)L(0,1) \\
&\quad + L'(0,1)L(0,0) + L'(0,0)L(0,1) \\
&\quad -2L'(0,1)L(0,1)+2L'(0,0)L(0,0) \big\} \ ,
\end{alignedat}
\raisetag{2\baselineskip}
\\
&H_{21}  = H_{12} \ ,
\\
&\begin{alignedat}{8}
H_{22}  = (&\mathbf{m}+\overline{\mathbf{m}})^2 + 3\kappa^2 -\dfrac{G_\pi b^2}{\pi}\overline{\mathbf{m}}\mathbf{m} \\
 \times \big\{ &L'(1/2,1/2)L(-1/2,-1/2) \\
  + &L'(-1/2,1/2)L(1/2,-1/2) \\
  +&L'(1/2,-1/2)L(-1/2,1/2) \\
  + &L'(-1/2,-1/2)L(1/2,1/2) \\
  +&L'(-1/2,3/2)L(-1/2,-1/2) \\
 -&2L'(-1/2,1/2)L(-1/2,1/2) \\
  + &L'(-1/2,-1/2)L(-1/2,3/2) \big\} \\
 -&\dfrac{2 G_\pi b^2}{\pi} \big\{ \overline{\mathbf{m}}\,L'(-1/2,1/2) \\
+ &\mathbf{m}\,L'(1/2,-1/2) \big\} \big\{\overline{\mathbf{m}}\,L(-1/2,1/2) \\
+& \mathbf{m}\,L(1/2,-1/2) \big\} \,
\end{alignedat}
\\
&\begin{alignedat}{8}
H_{23} & = \dfrac{2 G_\pi b^2}{\pi} [\overline{\mathbf{m}}\,L'(-1/2,1/2) + \mathbf{m}\,L'(1/2,-1/2) ] \\
& \quad \times [\overline{\mathbf{m}}\,L(-1/2,1/2)+\mathbf{m}\,L(1/2,-1/2) ] \ ,
\end{alignedat}
\\
&\begin{alignedat}{8}
H_{24} & = \dfrac{2 G_\pi b^3}{\pi} \overline{\mathbf{m}}\big\{  [L'(-1/2,3/2) +2L'(-1/2,1/2) ] \\
& \times L(0,0) - L'(-1/2,1/2)L(0,1)\big\} \\
& + \dfrac{2 G_\pi b^3}{\pi} \mathbf{m} \big\{ [L'(1/2,1/2)+2L'(1/2,-1/2)]\\
& \quad \quad \times L(0,0)  + L'(1/2,-1/2)L(0,1) \big\}\ ,
\end{alignedat}
\raisetag{3\baselineskip}
\\
&H_{31} = H_{13} \ ,
\\
&H_{32} = H_{23} \ ,
\\
&\begin{alignedat}{8}
&H_{33}  = (\mathbf{m}+\overline{\mathbf{m}})^2 + 3\kappa^2 -\dfrac{G_\pi b^2}{\pi}\overline{\mathbf{m}}\mathbf{m} \\
&\times \big\{ L'(1/2,1/2)L(-1/2,-1/2)  \\
&  + L'(-1/2,1/2)L(1/2,-1/2) \\
&  + L'(1/2,-1/2)L(-1/2,1/2) \\
&  + L'(-1/2,-1/2)L(1/2,1/2) \\
&  + L'(-1/2,3/2)L(-1/2,-1/2) \\
&  - 2L'(-1/2,1/2)L(-1/2,1/2) \\
&  + L'(-1/2,-1/2)L(-1/2,3/2) \big\} \\
& -\dfrac{2 G_\pi b^2}{\pi} [\overline{\mathbf{m}}\,L'(-1/2,1/2) + \mathbf{m}\,L'(1/2,-1/2) ]\\
&\hspace{1.233cm}
\times[ \overline{\mathbf{m}}\,L(-1/2,1/2)+\mathbf{m}\,L(1/2,-1/2)]\ ,
\end{alignedat}
\raisetag{7\baselineskip}
\\
&\begin{alignedat}{7}
H_{34}  = -\dfrac{4 G_\pi b^3}{\pi}&\bigg\{ \mathbf{m}\big\{L'(1/2,-1/2)L(0,1)  \\
&   + L'(1/2,-1/2)L(0,0)\big\}  \\
& + \overline{\mathbf{m}} \, L'(-1/2,1/2)L(0,0)\bigg\}\ ,
\end{alignedat}
\\
& H_{41} = H_{14} \ , \\
& H_{42} = H_{24} \ , \\
& H_{43} = H_{34} \ , \\
&
\begin{alignedat}{8}
H_{44} = (\mathbf{m}+\overline{\mathbf{m}})^2 &+ 5\kappa^2 \\&-\dfrac{8 G_\pi b^4}{\pi} L'(0,0)L(0,0) \ .
\end{alignedat}
\end{align}
\end{subequations}

\section{Analytical expressions for integrals of basis functions}
\subsection{Integrals for the calculation of the decay constant\label{ss:itg_decay_constant}}
When calculating the decay constants using the valence LFWFs of mesons, we encounter the following integral:
\begin{align}
\begin{alignedat}{7}
& \int_0^1\dfrac{\d{x}}{4\pi\sqrt{x(1-x)}}\int \dfrac{d^2\kappa^\perp}{(2\pi)^2}\,\phi_{nm}\left(\dfrac{\vv{\kappa}^\perp}{\sqrt{x(1-x)}} \right)\,\chi_l(x)\\
& =  \int_{0}^{2\pi}\dfrac{d\phi}{2\pi}\,e^{im\phi}\int_0^1\dfrac{\d{x}}{4\pi}\,\sqrt{x(1-x)}\chi_l(x)\,\int \dfrac{\rho \d\rho}{2\pi b}\\
& \quad \times \sqrt{\dfrac{4\pi n!}{(n+\vert m\vert)!}}\left(\dfrac{q}{b}\right)^{\vert m\vert} e^{-q^2/(2b^2)}L_n^{\vert m\vert }(q^2/b^2)q^{\vert m \vert}\\
& = \delta_{m,0}\,L_l(1/2,1/2;\alpha,\beta)\,\dfrac{b}{\sqrt{\pi}}(-1)^n \ ,\label{eq:itg_decay_constant}
\end{alignedat}
\raisetag{1.5\baselineskip}
\end{align}
where $L_l(1/2,1/2;\alpha,\beta)$ is given by Eq.~\eqref{eq:def_L1_ab_alphabeta}. We have also used
\begin{equation}
\int_{0}^{+\infty}\dfrac{q\d q}{2\pi b}\sqrt{4\pi} e^{-q^2/(2b^2)}L_n^0(q^2/b^2) =\dfrac{b}{\sqrt{\pi}}(-1)^n
\end{equation}
in deriving Eq.~\eqref{eq:itg_decay_constant}.

\subsection{Integrals for the mass radius\label{ss:itg_mass_radius}}
	To evaluate the transverse integrals in Eq.~\eqref{eq:def_I_m}, we first define $\vv{\rho}=b\sqrt{x(1-x)}\vv{r}^\perp$. After this substitution of variables  
	we obtain
\begin{align}
\begin{alignedat}{8}
& \quad \int_{0}^{+\infty}\d|\vv{r}^\perp|^2\int_{0}^{2\pi}\dfrac{\d\phi_r}{2}x^2(1-x)^2\,b^2|\vv{r}^\perp|^2 \\ & \quad \times \tilde{\phi}^*_{n',m'}\left(\sqrt{x(1-x)}\vv{r}^\perp\right)\tilde{\phi}_{n,m}\left(\sqrt{x(1-x)}\vv{r}^\perp\right) \\
& = b^{-2}\int_{0}^{+\infty}\d\rho^2 \int_{0}^{2\pi}\dfrac{\d\phi_r}{2}\rho^2 \tilde{\phi}^*_{n',m'}\left(\vv{\rho}^\perp/b\right) \\
& \quad \times \tilde{\phi}_{n,m} \left(\vv{\rho}^\perp/b\right) \\
& =(-1)^{n'+n}\delta_{m'm}\sqrt{\dfrac{n'!n!}{(n'+|m|)! (n+|m|)!}}  \\
& \quad \times \int_{0}^{+\infty}\d\rho^2\,(\rho^2)^{|m|+1}e^{-\rho^2}L_{n'}^{|m|}(\rho^2)L_{n}^{|m|}(\rho^2) \ .
\end{alignedat}
\raisetag{6\baselineskip}
\end{align}
The integrals over the product of generalized Laguerre polynomials can be obtained by the orthonormality relations. In order to apply such relations, we convert $L_n^{|m|}$ into $L_n^{|m|+1}$ using recurrence relations. From Eq.~(22.7.30) of Ref.~\cite{abramowitz1964handbook} we obtain
\begin{equation}
L_n^{|m|}(\rho^2)=L_n^{|m|+1}(\rho^2)-L_{n-1}^{|m|+1}(\rho^2) \ .
\end{equation}
Here when $n=0$, the second term drops out. We then have
\begin{align}
\begin{alignedat}{8}
& \sqrt{\dfrac{n'!n!}{(n'+|m|)! (n+|m|)!}}
\\
&\times \int_{0}^{+\infty}\d\rho^2\,(\rho^2)^{|m|+1}e^{-\rho^2}L_{n'}^{|m|}(\rho^2)L_{n}^{|m|}(\rho^2)
\\
=&\sqrt{\dfrac{n'!n!}{(n'+|m|)! (n+|m|)!}}\int_{0}^{+\infty}\d\rho^2(\rho^2)^{|m|+1}e^{-\rho^2}
\\
&\times\left[L_{n'}^{|m|+1}(\rho^2)-\theta_{n'-1}L_{n'-1}^{|m|+1}(\rho^2)\right]
\\
&\times \left[L_{n}^{|m|+1}(\rho^2)-\theta_{n-1}L_{n-1}^{|m|+1}(\rho^2)\right]
\\
=&\sqrt{\dfrac{n'!n!}{(n'+|m|)! (n+|m|)!}}\bigg\{\dfrac{(n+|m|+1)!}{n!}\delta_{n'n}
\\
& -\dfrac{(n+|m|)!}{(n-1)!}\delta_{n',n-1}\theta_{n-1}-\dfrac{(n+|m|+1)!}{n!}\delta_{n',n+1}
\\
& +\dfrac{(n+|m|)!}{(n-1)!}\delta_{n'n}\theta_{n-1} \bigg\}
\\
= &(2n+|m|+1)\delta_{n'n}-\sqrt{n(n+|m|)}\delta_{n',n-1}
\\
& -\sqrt{(n+1)(n+|m|+1)}\delta_{n',n+1} \ .
\end{alignedat}
\raisetag{10\baselineskip}
\end{align}
with $n\in\mathbf{N}$ by default.

The longitudinal integrals can be calculated by applying the orthonormal relation of the longitudinal basis function. Subsequently, we obtain the following expression for the hermitian matrix that specifies the mass radius:
\begin{align}
\begin{alignedat}{8}
& I_m(n',m',l';n,m,l) \\
& = \delta_{l'l}\,\delta_{m'm} \bigg\{ (2n+|m|+1)\delta_{n'n}  \\
&  +\sqrt{n(n+|m|)}\delta_{n',n-1} \\
&  +\sqrt{(n+1)(n+|m|+1)}\delta_{n',n+1} \bigg\} \ .
\end{alignedat}
\end{align}

\subsection{Longitudinal integrals
\label{ss:def_L_a_b}}	
	Let us define the following integral in the longitudinal basis functions:
	\begin{align}
	\label{eq:def_L1_ab_alphabeta}
	\begin{alignedat}{9}
	& L_l(a,b;\alpha,\beta) \\
	\equiv &\int_{0}^{1}\dfrac{\d{x}}{4\pi}\,x^b(1-x)^a\chi_l(x;\alpha,\beta)\\
	 =&\sqrt{\dfrac{2l+\alpha+\beta+1}{4\pi}}\sqrt{\dfrac{\Gamma(l+1)\Gamma(l+\alpha+\beta+1)}{\Gamma(l+\alpha+1)\Gamma(l+\beta+1)}}\\
	& \times \int_{0}^{1}\d{x}\,x^{\beta/2+b}(1-x)^{\alpha/2+a}\,P_l^{(\alpha,\beta)}(2x-1) \\
	=&\sqrt{\dfrac{2l+\alpha+\beta+1}{4\pi}}\sqrt{\dfrac{\Gamma(l+1)\Gamma(l+\alpha+\beta+1)}{\Gamma(l+\alpha+1)\Gamma(l+\beta+1)}}\\
	&\times\sum_{m=0}^{l}
	\begin{pmatrix}
	l+\alpha \\
	m
	\end{pmatrix}
	\begin{pmatrix}
	l+\beta \\
	l-m
	\end{pmatrix}(-1)^{l-m}\\
	& \times B\left(\dfrac{\beta}{2}+b+m+1,\dfrac{\alpha}{2}+a+l-m+1\right) ,
	\end{alignedat}
	\raisetag{3.5\baselineskip}
	\end{align}
	where $B(s,t)=\Gamma(s)\Gamma(t)/\Gamma(s+t)$ is the Euler Beta function.
	
	To evaluate $L_l(a,b;\alpha,\beta)$ numerically, we first rewrite Eq.~\eqref{eq:def_L1_ab_alphabeta} as
	\begin{equation}
	\begin{alignedat}{8}
	L_l(a,b;\alpha,\beta)=&\sqrt{\dfrac{2l+\alpha+\beta+1}{4\pi}}
	\\
	&\times\hspace{1cm}
	\sum_{m=0}^{l}C_{l,m}(a,b;\alpha,\beta) \ ,
	\end{alignedat}
	\end{equation}
	with
    \begin{align}
    \begin{alignedat}{8}
    C_{l,m}
	&\equiv\dfrac{(-1)^{l-m}\sqrt{\Gamma(l+1)\Gamma(l+\alpha+\beta+1)}}{\Gamma(m+1)\Gamma(l+\alpha-m+1)}\\
	& \times \dfrac{\sqrt{\Gamma(l+\alpha+1)\Gamma(l+\beta+1)}}{\Gamma(l-m+1)\Gamma(\beta+m+1)}\\
	& \times\dfrac{\Gamma(\beta/2+b+m+1)\Gamma(\alpha/2+a+l-m+1)}{\Gamma(\beta/2+b+\alpha/2+a+l+2)} \ .
    \end{alignedat}
    \raisetag{3.5\baselineskip}
	\end{align}
	We then obtain the following recurrence relations for $C_{l,m}$:
	\begin{subequations}
	\label{eq:C_lm_recurrence}
    \begin{align}
    &\begin{alignedat}{8}
    C_{0,0} & =\sqrt{\dfrac{\Gamma(\alpha+\beta+1)}{\Gamma(\alpha+1)\Gamma(\beta+1)}} \\
	&\times \dfrac{\Gamma(\beta/2+b+1)\Gamma(\alpha/2+a+1)}{\Gamma(\beta/2+b+\alpha/2+a+2)} \ ,
    \end{alignedat}
    \\
    &\begin{alignedat}{8}
    \dfrac{C_{l,0}}{C_{l-1,0}}& =-\sqrt{\dfrac{(l+\beta)(l+\alpha+\beta)}{l(l+\alpha)}}\\
	& \times \dfrac{\alpha/2+a+l}{\beta/2+b+\alpha/2+a+l+1} \quad (\text{for } l\geq 1)\ ,
    \end{alignedat}
    \raisetag{2.5\baselineskip}
    \\
    &\begin{alignedat}{8}
    \dfrac{C_{l,m}}{C_{l,m-1}} & =
	-\dfrac{(l+\alpha-m+1)(l-m+1)}{m(\beta+m)(\alpha/2+a+l-m+1)}
	\\
	& \times (\beta/2+b+m)	\quad (\text{for } l\geq m\geq 1) \ .
    \end{alignedat}
    \raisetag{1.5\baselineskip}
    \end{align}
	\end{subequations}
	The longitudinal integral $L_l(a,b;\alpha,\beta)$ can then be calculated by first generating and then summing the following sequences:
	\begin{equation*}
	\begin{array}{l}
	C_{0,0} \\
	\downarrow \\
	C_{1,0}+C_{1,1} \\
	\downarrow \\
	C_{2,0}+C_{2,1}+C_{2,2} \\
	\downarrow \\
	C_{3,0}+C_{3,1}+C_{3,2}+C_{3,3} \\
	\downarrow \\
	\dots\,\dots
	\end{array},
	\end{equation*}
	using Eq.~\eqref{eq:C_lm_recurrence}.
	
	\section{The electromagnetic form factors in the basis representation}
	\subsection{Reduction of the formula for the electromagnetic form factors in the valence Fock sector of mesons\label{eq:red_Drell_Yan_West}}
	Let us first expand the quark current operator in terms of creation and annihilation operators. In agreement with the light-front quantization condition, the Dirac field operator at a given light-front time $x^+=0$ is expanded according to
	\begin{equation}
	\begin{alignedat}{8}
	\psi&(x)  =\!\!\!\sum_{s=\pm 1/2} \int \d{}\underline{p}
	\\
	&
	\times \left[ b_s(p)u_s(p)e^{-ip\cdot x} + d^\dagger_s(p)v_{s}(p)e^{ip\cdot x}\right]
	\mathrlap{
	\bigg\vert_{x^+=0} \ ,
	}
	\end{alignedat}
	\end{equation}
	where the flavor indices are implicit. Here $u_s(p)$ and $v_s(p)$ are solutions of the Dirac equation for free fermions. Meanwhile, the creation and annihilation operators satisfy these anti-commutation relations:
	\begin{align}
	\{ b_r(k),\,b_s^\dagger(p) \} & = \underline{\delta}(k-p) \delta_{rs} \ ,\\
	\{ d_r(k),\,d_s^\dagger(p) \} & = \underline{\delta}(k-p) \delta_{rs} \ ,
	\end{align}
	while other anti-commutation relations all vanish. We have defined the integral measure in the momentum space as
	\begin{equation}
	\int \d{}\underline{p} = \int_{0}^{+\infty}\dfrac{\d p^+}{4\pi p^+}\int_{-\infty}^{+\infty}\dfrac{\d p^\perp_1}{2\pi}\int_{-\infty}^{+\infty}\dfrac{\d p^\perp_2}{2\pi} \ .
	\end{equation}
	The reduced delta-function is defined according to
	\begin{equation}
	\begin{alignedat}{8}
	\underline{\delta}(k-p)= 4\pi k^+\theta(k^+)\,&\delta\bigl(k^+-p^+\bigr)
	\\\times
	(2\pi)^2&\delta\bigl(\vv{k}^\perp -\vv{p}^\perp \bigr) \ .
	\end{alignedat}
	\end{equation}
	These conventions ensure that one reduced delta-function can be utilized to eliminate one momentum-space integration.
	
	With these definitions, the charge density operator becomes
	\begin{align}
	\begin{alignedat}{8}
	& \quad \lim\limits_{x\rightarrow 0}\mathrm{e}_{\mathrm{f}}\,\psi(x)\gamma^+ \psi(x)
	= \lim\limits_{x\rightarrow 0}\sum_{s's} \int \d{}{}\underline{p}' \int \d{}{}\underline{p}\,\\
	& \quad \times \left[b^\dagger_{s'}(p')\overline{u}_{s'}(p')e^{ip'\cdot x}+d_{s'}(p')\overline{v}_{s'}(p')e^{-ip\cdot x} \right]\\
	& \quad \times \mathrm{e}_{\mathrm{f}}\gamma^+\left[b_s(p)u_{s}(p)e^{-ip\cdot x}+ d^\dagger_s(p)v_s(p)e^{ip\cdot x} \right]\\
	& \rightarrow \sum_{s's} \int \d{}{}\underline{p}'\int \d{}\underline{p} \bigg\{\mathrm{e}_{\mathrm{q}}\,b^\dagger_{s'}(p')b_s(p)\,\overline{u}_{s'}(p')\gamma^+ u_s(p)\\
	&\quad -\mathrm{e}_{\overline{\mathrm{q}}}\,d^\dagger_{s'}(p')d_s(p)\,\overline{v}_{s'}(p')\gamma^+ v_s(p) \bigg\} \\
	& = \sum_s \int \d{}\underline{p}'\int \d{}\underline{p}\,2\sqrt{p'^+p^+}\\
	& \quad \quad \left[\mathrm{e}_{\mathrm{q}}\,b^\dagger_s(p')b_s(p)-\mathrm{e}_{\overline{\mathrm{q}}}\,d^\dagger_s(p')d_s(p) \right] \ .
    \end{alignedat}
    \raisetag{\baselineskip}
    \end{align}
	Here we have made use of ${\overline{u}_{s'}(p')\gamma^+ u_s(p)} = {2\sqrt{p'^+ p^+}\delta_{s's}}$ and ${\overline{v}_{s'}(p')\gamma^+ v_s(p)} = {2\sqrt{p'^+ p^+}\delta_{s's}}$.
	We have only kept terms of relevance to the valence Fock sector of mesons.
	The form factors then becomes
	\begin{align}
	\begin{alignedat}{7}
	& \,\,\, I_{m'_J,m_J}(Q^2) \\
	&
	\mathrlap{
	=\sum_{r's'}\int_{0}^{1}\dfrac{\d{x}'}{4\pi x'(1-x')}\int \dfrac{\d{}\vv{\kappa}'^\perp}{(2\pi)^2}\,\psi_{r's'}^*(x',\vv{\kappa}'^\perp )
	}
	\\
	& \times \langle 0 \vert d_{s'}(k_2')b_{r'}(k_1') \sum_\sigma \int \d{}\underline{p}'\int \d{}\underline{p} \\
	&  \times \dfrac{\sqrt{p'^+p^+}}{P^+} \left[\mathrm{e}_{\mathrm{q}}\,b^\dagger_\sigma(p')b_\sigma(p)-\mathrm{e}_{\overline{\mathrm{q}}}\,d^\dagger_\sigma(p')d_\sigma(p) \right] \\
	& \times \sum_{rs}\int_{0}^{1}\dfrac{\d{x}}{4\pi x(1-x)}\int \dfrac{\d{}\vv{\kappa}^\perp}{(2\pi)^2} \\
	&  \times b^\dagger_r(k_1)d^\dagger_s(k_2)\vert 0\rangle \,\psi_{rs}(x,\vv{\kappa}^\perp) \ .
	\end{alignedat}
	\end{align}
	The anti-commutation relation for the 
	creation and annihilation operator can be used to deduce
	\begin{align}
	\begin{multlined}
	\sum_\sigma \int \d{}\underline{p}'\int \d{}\underline{p}\,d_{s'}(k_2')b_{r'}(k_1')b^\dagger_\sigma(p')b_\sigma(p)   \\
	 \times b^\dagger_r(k_1)d^\dagger_s(k)\sqrt{p'^+p^+} \rightarrow
	 \sqrt{k_1'^+ k_1^+}\delta_{r'r}\delta_{s's}\underline{\delta}
	 (k_2'-k_2) \ ,
	\end{multlined}
	\raisetag{2.5\baselineskip}
	\\
	\begin{multlined}
	\sum_\sigma \int \d{}\underline{p}'\int \d{}\underline{p}\,d_{s'}(k_2')b_{r'}(k_1')d^\dagger_\sigma(p')d_\sigma(p) \\
	\times b^\dagger_r(k_1)d^\dagger_s(k)\sqrt{p'^+p^+} \rightarrow \sqrt{k_2'^+ k_2^+}\delta_{r'r}\delta_{s's}\underline{\delta}(k_1'-k_1) \ .
	\end{multlined}
	\raisetag{3.5\baselineskip}
	\end{align}
	The expression for the form factors is then reduced to
	\begin{align}
	\begin{alignedat}{8}
	& I_{m'_J,m_J}(Q^2) \\
	& = \sum_{rs}\int \dfrac{\d{x}'}{4\pi x'(1-x')} \int \dfrac{\d{}\vv{\kappa}^\perp}{(2\pi)^2}\,\psi^*_{rs}(x',\vv{\kappa}'^\perp)\\
	& \times \int \dfrac{\d{x}}{4\pi x(1-x)}\int \dfrac{\d{}\vv{\kappa}^\perp}{(2\pi)^2}\psi_{rs}(x,\vv{\kappa}^\perp) \Bigg\{ \mathrm{e}_{\mathrm{q}} \\
	& \times
	\mathrlap{
	\dfrac{\sqrt{k_1'^+ k_1^+}}{P^+}\,\underline{\delta}(k_2'-k_2) - \mathrm{e}_{\overline{\mathrm{q}}}\dfrac{\sqrt{k_2'^+k_2^+}}{P^+}\underline{\delta}(k_1'-k_1) \Bigg\} \ ,
	}
	\end{alignedat}
	\end{align}
	where we have defined
	\begin{subequations}
	\begin{align}
	k_1^+ &= x P^+ \ , \\
	\vv{k}_1^\perp &= \vv{\kappa}^\perp + x \vv{P}^\perp \ , \\
	k_2^+ &= (1-x) P^+ \ , \\
	\vv{k}_2^\perp &= -\vv{\kappa}^\perp + (1-x) \vv{P}^\perp \ ,
	\intertext{and}
	k_1' &= x' P'^+ \ , \\
	\vv{k}_1'^\perp &= \vv{\kappa}' + x' \vv{P}'^\perp \, \\
	k_2'^+ &= (1-x') P'^+ \ , \\
	\vv{k}_2'^\perp &= -\vv{\kappa}'^\perp + (1-x') \vv{P}'^\perp \ .
	\end{align}
	\end{subequations}
	Meanwhile, the following reductions of delta-functions hold in the Drell-Yan frame:
	\begin{align}
	&\begin{alignedat}{8}
	& \!\!\!\!\!\!\underline{\delta}(k_2'-k_2) = 4\pi (1-x)\delta(x'-x)\,(2\pi)^2 \\
	& \times \delta^2\left(-\vv{\kappa}'^\perp +\vv{\kappa}^\perp +(1-x)(\vv{P}'^\perp -\vv{P}^\perp) \right) \ ,
	\end{alignedat}
	\raisetag{2\baselineskip}
	\\
	&\begin{alignedat}{8}
	& \!\!\!\!\!\!\underline{\delta}(k_1'-k_1) = 4\pi x\,\delta(x'-x)\,(2\pi)^2\\
	& \times \delta^2\left(\vv{\kappa}'^\perp -\vv{\kappa}^\perp +x(\vv{P}'^\perp -\vv{P}^\perp ) \right) \ .
	\end{alignedat}
	\end{align}
	Therefore the expression for the form factors is reduced to that given on the right-hand side of Eq.~\eqref{eq:def_I_mj,mj'} with $\vv{q}^\perp=\vv{P}'^\perp -\vv{P}^\perp$.
	\subsection{The electromagnetic form factors in the basis representation\label{ss:Drell_Yan_West_basis_rep}}
	In the basis representation, Eq.~\eqref{eq:def_I_mj,mj'} becomes
	\begin{align}
	\begin{alignedat}{8}
	&\quad I_{m_J',m_J}(Q^2)\\
	& = \sum_{n'm'l'} \sum_{nml} \sum_{rs} \,\psi^* _{n'm'l'rs}\,\psi_{nmlrs} \\
	& \quad \times \int \dfrac{\d{x}}{4\pi x(1-x)} \chi_{l'}(x)\chi_l(x) \int \dfrac{\d{}\vv{k}}{(2\pi)^2} \\
	& \quad \times \Bigg\{\mathrm{e}_{\mathrm{q}}\, \phi_{n'm'}^*\left(\dfrac{\vv{k}^\perp +(1-x)\vv{q}^\perp}{\sqrt{x(1-x)}} \right)  \\
	&
	\mathrlap{
	\quad \quad - \mathrm{e}_{\overline{\mathrm{q}}}\,\phi_{n'm'}^*\left(\dfrac{\vv{k}^\perp - x\vv{q}^\perp}{\sqrt{x(1-x)}} \right) \Bigg\} \phi_{nm}\left(x,\vv{k}^\perp\right)
	}
	\\
	& = \sum_{n'm'l'} \sum_{nml} \sum_{rs} \,\psi^* _{n'm'l'rs}\,\psi_{nmlrs} \\
	& \quad \times \int \dfrac{\d{x}}{4\pi x(1-x)} \chi_{l'}(x)\chi_l(x) \int \dfrac{\d{}\vv{k}}{(2\pi)^2} \\
	& \quad \times \Bigg\{\mathrm{e}_{\mathrm{q}}\,\phi_{n',-m'}\left(\dfrac{\vv{k}^\perp +(1-x)\vv{q}^\perp/2}{\sqrt{x(1-x)}} \right) \\
	& \quad \hspace{1.5cm} \times \phi_{nm}\left(\dfrac{\vv{k}^\perp -(1-x)\vv{q}^\perp/2}{\sqrt{x(1-x)}} \right) \\
	&\quad\quad -  \mathrm{e}_{\overline{\mathrm{q}}}\,\phi_{n',-m'}\left(\dfrac{\vv{k}^\perp - x\vv{q}^\perp/2}{\sqrt{x(1-x)}} \right) \\
	&\quad \hspace{1.5cm} \times \phi_{nm}\left(\dfrac{\vv{k}^\perp+x\vv{q}^\perp/2}{\sqrt{x(1-x)}} \right) \Bigg\} \ ,
	\end{alignedat}
	\end{align}
	where we have applied shifts in the transverse momentum and $\phi_{nm}^*=\phi_{n,-m}$.
	
	We then apply the Talmi-Moshinsky (TM) transform to simplify the integrals in the transverse momentum~\cite{positronium,qua10800}. Specifically, the we have
	\begin{align}
	\begin{multlined}
	\phi_{n',-m'}(\vv{q}_1)\,\phi_{n,m}(\vv{q}_2) \\
	 = \sum_{NM\overline{n}\overline{m}} C(n',-m',n,m;N,M,\overline{n},\overline{m})\,\\
	 \times \phi_{NM}(\vv{P})\,\phi_{\overline{n}\overline{m}}(\vv{p}) \ ,
	\end{multlined}
	\end{align}
	with all $4$ harmonic oscillator functions sharing the same scale $b$ and
	\begin{equation}
	\begin{cases}
	\vv{P} = (\vv{q}_1+\vv{q}_2)/\sqrt{2} \\
	\vv{p} = (\vv{q}_1-\vv{q}_2)/\sqrt{2}
	\end{cases},
	\end{equation}
	which corresponds to
	\begin{equation}
	\begin{cases}
	\vv{P} = \dfrac{\sqrt{2}\vv{k}^\perp}{\sqrt{x(1-x)}}\\
	\vv{p} = \sqrt{\dfrac{1-x}{2x}} \vv{q}^\perp
	\end{cases}
	\end{equation}
	for the quark contribution and
	\begin{equation}
	\begin{cases}
	\vv{P} = \dfrac{\sqrt{2}\vv{k}^\perp}{\sqrt{x(1-x)}}\\
	\vv{p} = -\sqrt{\dfrac{x}{2(1-x)}} \vv{q}^\perp
	\end{cases}
	\end{equation}
	for the anti-quark contribution.
	
	The coefficient $C(n',-m',n,m;N,M,\overline{n},\overline{m})$ can be computed with established procedures~\cite{positronium,qua10800}. The following observations made specifically for the valence Fock sector of mesons will be helpful in enumerating terms after in the TM transform.
	\begin{itemize}
		\item Because the TM transform cannot change the total magnetic projection of the orbital angular momentum, we must have $-m'+m=M+\overline{m}$.
		\item The integral in $\vv{k}^\perp$ will select the terms with $\overline{m}=0$, leaving other values of $m$ not contributing to the integral.
		\item Because the mesons obtained from the light-front Hamiltonian have fixed magnetic projection for the sum of the spin and orbital angular momenta, when the spins of the two wave-function in a bilinear are identical, so are their magnetic quantum numbers $m'$ and $m$.
	\end{itemize}
	These observations further confine us to $\overline{m}=-m'+m=0$, which is expected since the electromagnetic form factors have no angular dependence.
	
	The integrals over the momenta of the light-front wave functions then become
	\begin{align}
	& \quad \tilde{C}(n',m',l';n,m,l;Q^2) \nonumber\\
	& \equiv \int \dfrac{\d{x}}{4\pi x(1-x)}\chi_{l'}(x)\chi_l(x)\int \dfrac{\d{} \vv{k}^\perp}{(2\pi)^2} \nonumber\\
	& \quad \times \Bigg\{\mathrm{e}_{\mathrm{q}}\, \phi_{n',-m'}\left(\dfrac{\vv{k}^\perp +(1-x)\vv{q}^\perp/2}{\sqrt{x(1-x)}} \right) \nonumber\\
	& \quad \quad \hspace{2cm} \times
	\phi_{nm}\left(\dfrac{\vv{k}^\perp -(1-x)\vv{q}^\perp/2}{\sqrt{x(1-x)}} \right) \nonumber\\
	&\quad \quad -  \mathrm{e}_{\overline{\mathrm{q}}}\,\phi_{n',-m'}\left(\dfrac{\vv{k}^\perp - x\vv{q}^\perp/2}{\sqrt{x(1-x)}} \right) \, \nonumber\\
	& \quad \quad \hspace{2cm} \times  \phi_{nm}\left(\dfrac{\vv{k}^\perp+x\vv{q}^\perp/2}{\sqrt{x(1-x)}} \right) \Bigg\} \nonumber \\
	& = \int \dfrac{\d{x}}{4\pi x(1-x)}\chi_{l'}(x)\chi_l(x)\nonumber\\
	& \quad \times \sum_{NM\overline{n}\overline{m}}C(n',-m',n,m;N,M,\overline{n},\overline{m})\, \int \dfrac{\d{}\vv{k}}{(2\pi)^2} \nonumber\\
	& \quad \times \phi_{NM}\left(\dfrac{\sqrt{2}\vv{k}^\perp}{\sqrt{x(1-x)}} \right) \Bigg\{ \mathrm{e}_{\mathrm{q}}\,\phi_{\overline{n}\overline{m}}\left(\sqrt{\dfrac{1-x}{2x}}\vv{q}^\perp \right) \nonumber\\
	& \quad \quad \hspace{2cm} - \mathrm{e}_{\overline{\mathrm{q}}}\,\phi_{\overline{n}\overline{m}}\left(-\sqrt{\dfrac{x}{2(1-x)}}\vv{q}^\perp \right) \Bigg\}\nonumber\\
	& = \int \dfrac{\d{x}}{4\pi}\chi_{l'}(x)\chi_l(x) \sum_{NM \overline{n}  \overline{m}} \dfrac{(-1)^N b}{2\sqrt{\pi}}\delta_{M0} \nonumber\\
	& \quad \times C(n',-m',n,m;N,M,\overline{n},\overline{m})\, \nonumber \\
	& \quad \times \Bigg\{ \mathrm{e}_{\mathrm{q}}\,\phi_{\overline{n}\overline{m}}\left(\sqrt{\dfrac{1-x}{2x}}\vv{q}^\perp \right) \nonumber\\
	& \quad \hspace{2cm} - \mathrm{e}_{\overline{\mathrm{q}}}\,\phi_{\overline{n}\overline{m}}\left(-\sqrt{\dfrac{x}{2(1-x)}}\vv{q}^\perp \right) \Bigg\}\nonumber \\
	& = \delta_{m'm}\sum_{N\overline{n}}C(n',-m',n,m;N,0,\overline{n},0) \nonumber\\
	& \quad \times \int \dfrac{\d{x}}{4\pi}\chi_{l'}(x)\chi_l(x)\, \dfrac{(-1)^N b}{2\sqrt{\pi}} \nonumber \\
	&\quad \times \Bigg\{ \mathrm{e}_{\mathrm{q}}\,\phi_{\overline{n}0}\left(\sqrt{\dfrac{1-x}{2x}}\vv{q}^\perp \right) \nonumber \\
	& \quad \hspace{2cm} - \mathrm{e}_{\overline{\mathrm{q}}}\,\phi_{\overline{n}0}\left(-\sqrt{\dfrac{x}{2(1-x)}}\vv{q}^\perp \right) \Bigg\}\nonumber
	\\&= \delta_{m'm}\sum_{N\overline{n}}C(n',-m',n,m;N,0,\overline{n},0) \nonumber\\
	& \quad \times \int \dfrac{\d{x}}{4\pi}\chi_{l'}(x)\chi_l(x) \,(-1)^N \nonumber \\
	& \quad \times \Bigg\{ \mathrm{e}_{\mathrm{q}}\,\exp\left(-\dfrac{1-x}{2x}\dfrac{Q^{2}}{2b^2} \right)\,L_{\overline{n}}\left(\dfrac{1-x}{2x}\dfrac{Q^{2}}{b^2} \right) \nonumber\\
	& \quad - \mathrm{e}_{\overline{\mathrm{q}}}\,\exp\left(-\dfrac{x}{2(1-x)}\dfrac{Q^2}{2b^2} \right)\,L_{\overline{n}}\left(\dfrac{x}{2(1-x)}\dfrac{Q^2}{b^2} \right) \Bigg\} \ . \label{eq:def_c_tilde_DYW_basis_rep}
	\end{align}
	
	With the aid 
	of the TM transform, the electromagnetic form factors in the basis representation becomes
	\begin{align}
	\begin{alignedat}{7}
	I_{m_J',m_J}&(Q^2)  = \sum_{n'm'l'} \sum_{n,m,l} \sum_{r,s}\,\psi^*_{n'm'l'rs}\\
	& \times \tilde{C}(n',m',l';n,m,l;Q^2)\,\psi_{nmlrs} \ .
	\end{alignedat}
	\end{align}

\section{Multi-qubit observables for the \texorpdfstring{${J_z=0}$}{Jz=0} sector of the \texorpdfstring{${N_{\mathrm{max}}=L_{\mathrm{max}}=0}$}{Nmax=Lmax=0}, \texorpdfstring{${M_{\mathrm{max}}=2}$}{Mmax=2} Hamiltonian\label{app:num_observ}}

Below we provide the multi-qubit expressions for observables discussed in Sec.~\ref{sec:observ} for the ${J_z=0}$ sector of the BLFQ pion Hamiltonian.

The decay constant can be obtained from eq.~\eqref{eq:fpa} as
\begin{equation}
\label{eq:subst_decay_const}
\begin{gathered}
        f_{\pi} = 61.6 \bigl|\sbraket{v}{\psi(\vv{\theta})}\bigr| \ ,\\
    \sket{v} = (0,1/\sqrt{2},-1/\sqrt{2},0) \ .
\end{gathered}
\end{equation}

In order to reduce the gate count, rather than using the circuit from Fig.~\ref{fig:linobservmeas}, we expand the projector onto the $\sket{v}$ state in terms of Pauli operators:
\begin{gather}
    \label{eq:subst_decay_const_proj}
    \saxe{v}{v} =
    \dfrac{1}{2}
    \begin{pmatrix}
    0 & \phantom{-}0 & \phantom{-}0 & \phantom{-}0 \\
    0 & \phantom{-}1 & -1           & \phantom{-}0 \\
    0 & -1           & \phantom{-}1 & \phantom{-}0 \\
    0 & \phantom{-}0 & \phantom{-}0 & \phantom{-}0
    \end{pmatrix}
    \ , \\
    \label{eq:subst_decay_const_direct}
    \begin{multlined}
    \saxe{v}{v}_{\text{direct}} =
    0.5IIII
    \\
    - 0.25 ( IXXI + IYYI + IZII + IIZI ) \ ,
    \end{multlined}
    \\
    \label{eq:subst_decay_const_compact}
    \saxe{v}{v}_{\text{compact}} =
    0.25 ( II - XX - YY - ZZ ) \ .
\end{gather}
\begin{sloppypar}
and calculate the decay constant using ${
    \bigl| \sbraket{v}{\psi(\vv{\theta})} \bigr| = \sqrt{ \bigl\langle\psi(\vv{\theta})\bigl|\bigl(\saxe{v}{v}\bigr)\bigr|\psi(\vv{\theta})\bigr\rangle }}
$.
\end{sloppypar}

The mass radius matrix is given by eq.~\eqref{eq:rm2_mass_radius}, which when expressed in terms of qubit operators is:
\begin{align}
\label{eq:subst_mass_rad_operator}
    &\dfrac{3}{2b^2}
    I_{m} =
    \begin{pmatrix}
    2.27 & 0 & 0 & 0 \\
    0 & 1.13 & 0 & 0 \\
    0 & 0 & 1.13 & 0 \\
    0 & 0 & 0 & 2.27
    \end{pmatrix}
    \ , \\
\label{eq:subst_mass_rad_direct}
&\begin{multlined}
\dfrac{3}{2b^2}I_{m,\,\text{direct}} =
- 1.30 ( IIIZ + ZIII ) \\
- 0.65 ( IIZI + IZII ) +
    3.92 IIII
     \ ,
\end{multlined}
    \\
\label{eq:subst_mass_rad_compact}
    &\dfrac{3}{2b^2}I_{m,\,\text{compact}} =
    1.96 II + 0.65 ZZ \ .
\end{align}

The $\rho_{l=0,l^\prime=0}$ density matrix of the parton distribution function is given by \mbox{eqs.~\eqref{eq:valence_PDF_lfwf}-\eqref{eq:rhopdf}}, which when expressed in terms of qubit operators is:
\begin{alignat}{8}
\label{eq:subst_pdf_rho_operator}
&\hspace{2cm}
    \rho =
    \begin{pmatrix}
    1 & 0 & 0 & 0 \\
    0 & 1 & 0 & 0 \\
    0 & 0 & 1 & 0 \\
    0 & 0 & 0 & 1
    \end{pmatrix} \ ,
    \\
    \label{eq:subst_pdf_rho_direct}
    &\begin{alignedat}{8}
    \rho_{\text{direct}} = \
    2 IIII-0.5 ( ZIII &+ IZII
    \\
    + IIZI &+ IIIZ ) \ ,
    \end{alignedat}
     \\
\label{eq:subst_pdf_rho_compact}
    &\rho_{\text{compact}} = II \ .
\end{alignat}

We calculate the elastic form factor matrix $F_{\mathrm{P}}(Q^2) $, eq.~\eqref{eq:def_F_pseudoscalar}, by discretizing $Q^2$ on the interval ${0\leq Q^2\leq5152900}$, evaluating the matrix $\tilde{C}(n',m',l';n,m,l;Q^2)$ for each value of $Q^2$, and expanding it in terms of Pauli operators.
For the sake of brevity, we do not include these explicit expressions for each point.

\section{Bravyi-Kitaev encoding\label{app:BK}}

Both Jordan-Wigner and Bravyi-Kitaev encoding allow one to store the second-quantized fermionic states in a quantum computer. Within the Jordan-Wigner encoding, each qubit stores the occupancy of a particular orbital~\cite{jordanwigner}. Within the Bravyi-Kitaev encoding, the information about parity is distributed equally between the operators and states~\cite{bravyi2002fermionic}.
In practice, one typically uses the more efficient BK encoding. While a single fermionic orbital in BK encoding is represented by up to $O(\log N)$ qubits (instead of $O(N)$ in JW), the creation and annihilation operators are represented now by $\log N$-local multi-qubit operators~\cite{bravyi2002fermionic}.

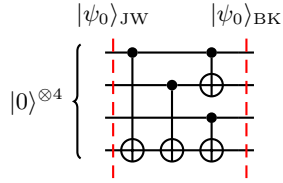
\begin{figure}[htp]
\centering
\begin{tikzpicture}
\node[scale=1]{
\begin{quantikz}[slice label style ={inner sep=10pt,anchor=south west,rotate=40}, column sep = .2cm, row sep = .2cm]
\lstick[wires=4]{$\sket{0}^{\otimes 4}$}
& \slice{$\sket{\psi_0}_{\text{JW}}$}& \ctrl{3} & \qw & \ctrl{1} & \qw \slice{$\sket{\psi_0}_{\text{BK}}$} & \qw \\
& & \qw & \ctrl{2} & \targ{} & \qw & \qw \\
& & \qw & \qw & \ctrl{1} & \qw & \qw \\
& & \targ{} & \targ{} & \targ{} & \qw & \qw
\end{quantikz}
};
\end{tikzpicture}
\caption{
Converting a four-qubit state from Jordan-Wigner to Bravyi-Kitaev encoding.
}
\label{fig:JWBKconvert}
\end{figure}

The BK-encoded basis states $\sket{\ldots b_2b_1b_0}$ can be obtained from the JW-encoded states by means of the linear transformation ${b_i = \sum_{ij} \mathcal{P}_{ij} f_j}$~\cite{BK2015}, where the entries of $\mathcal{P}_{ij}$ are $\{0,1\}$, and multiplicaation modulo 2 is implied. Such a transformation of states can be implemented efficiently on a quantum computer. Since the matrix $\mathcal{P}_{ij}$ is lower-triangular~\cite{BK2015}, multiplication modulo 2 can be performed on qubits using the CNOT gates, starting from the \emph{bottom} row. For example, in the case of four qubits, the encoded matrix has the form of
\begin{equation}
    \label{eq:JWBK_encoder}
    \mathcal{P}_{ij} =
    \begin{pmatrix}
    1 & 0 & 0 & 0 \\
    1 & 1 & 0 & 0 \\
    0 & 0 & 1 & 0 \\
    1 & 1 & 1 & 1
    \end{pmatrix} \ .
\end{equation}
The corresponding circuit is shown on Fig.~\ref{fig:JWBKconvert}.

In order to perform the simulation in the BK encoding, one adjusts the procedure outlined in Sec.~\ref{sec:direct} as follows: a) After preparing the JW-encoded initial state $\sket{\psi_{0}}$, one appends to the circuit a block converting JW-encoded states to BK-encoded ones as on Fig.~\ref{fig:JWBKconvert}; b) Eq.~\eqref{eq:JWmapterm} is replaced with its BK version which changes the coefficients $\alpha_j$ in~\eqref{eq:UCCevol} and $h_i$ in eq.~\eqref{eq:HPauliexp}.

\bibliography{main}
\bibliographystyle{unsrt}

\end{document}